\documentclass[a4paper,11pt]{article}

\usepackage[english]{babel}
\usepackage[utf8x]{inputenc}
\usepackage[T1]{fontenc}
\usepackage{bbm}
\usepackage{mathrsfs}
\usepackage{amssymb}
\usepackage{multirow}
\usepackage{bm}
\usepackage{longtable}
\usepackage{booktabs}
\usepackage{lscape}
\usepackage{caption}

\usepackage[a4paper,top=3cm,bottom=2cm,left=3cm,right=3cm,marginparwidth=1.75cm]{geometry}

\usepackage{amsmath}
\usepackage{graphicx}
\usepackage[colorinlistoftodos]{todonotes}
\usepackage[colorlinks=true, allcolors=blue]{hyperref}

\newtheorem{theorem}{Theorem}[section]
\newtheorem{lemma}[theorem]{Lemma}

\newenvironment{proof}[1][Proof]{\begin{trivlist}
		\item[\hskip \labelsep {\bfseries #1}]}{\end{trivlist}}

\newenvironment{remark}[1][Remark]{\begin{trivlist}
		\item[\hskip \labelsep {\bfseries #1}]}{\end{trivlist}}

\newcommand{\TCF}{\mathrm{TCF}}
\newcommand{\ROC}{\mathrm{ROC}}
\newcommand{\VUS}{\mathrm{VUS}}

\renewcommand{\P}{\mathrm{P}}

\newcommand{\N}{\mathrm{N}}

\newcommand{\Dir}{\mathrm{Dir}}
\newcommand{\Mult}{\mathrm{Mult}}
\newcommand{\TN}{\mathrm{TN}}

\title{Bayesian ROC surface estimation under verification bias}
\author{Rui Zhu and Subhashis Ghosal\\
North Carolina State University}

\begin{document}
\maketitle

\begin{abstract}
The Receiver Operating Characteristic (ROC) surface is a generalization of ROC curve and is widely used for assessment of the accuracy of diagnostic tests on three categories. A complication called the verification bias, meaning that not all subjects have their true disease status verified often occur in real application of ROC analysis. This is a common problem since the gold standard test, which is used to generate true disease status, can be invasive and expensive. In this paper, we will propose a Bayesian approach for estimating the ROC surface based on continuous data under a semi-parametric trinormality assumption. Our proposed method often adopted in ROC analysis can also be extended to situation in the presence of  verification bias. We compute the posterior distribution of the parameters under trinormality assumption by using a rank-based likelihood. Consistency of the posterior under mild conditions is also established. We compare our method with the existing methods for estimating ROC surface and conclude that our method performs well in terms of accuracy.

\noindent
\textbf{Key words}: ROC surface; Verification bias correction; Trinormal model; MAR assumption.
\end{abstract}

\section{Introduction}

The Receiver Operating Characteristic (ROC) curve analysis has been widely used as an effective tool in measuring the accuracy of diagnostic tests in the two-class classification problem. It shows a tradeoff between sensitivity and specificity by varying the cut-off point through all possible values of the diagnostic marker. 
The ROC surface is a generalization of the ROC curve which is intended to accommodate the problems of three-class classification (Scurfield, 1996). Let $X_1, X_2$ and $X_3$ be continuous measurements from three different classes, $X_1 \sim F_1$ are measurements from Class 1, $X_2 \sim F_2$ are measurements from Class 2, and $X_3 \sim F_3$ are measurements from Class 3.
Suppose that the ordering of interest for these three classes is $X_1<X_2<X_3$. A decision rule that classifies subjects can be defined by using two ordered threshold points $c_1<c_2$, i.e. choose Class 1 when a measurement is less than $c_1$, choose Class 2 when it is between $c_1$ and $c_2$, and choose Class 3 otherwise. This will result in three True Class Fractions (TCFs). 
\begin{align*}
\TCF_1 &=\P(X_1 \leq c_1)=F_1(c_1),\\
\TCF_2 &=\P(c_1 \leq X_2 \leq c_2)=F_2(c_2)-F_2(c_1),\\ \TCF_3 &=\P(X_3 > c_2)=1-F_3(c_2).
\end{align*}
Varying $c_1$ and $c_2$ will give us a set of TCFs. To construct the ROC surface, plot $(\TCF_1, \TCF_2, \TCF_3)$ in a three-dimensional coordinate system. The functional form of the ROC surface can be obtained by expressing $\TCF_2$ as a function of $(\TCF_1, \TCF_3)$ given by $\ROC_s(\TCF_1, \TCF_3)=F_2(F_3^{-1}(1-\TCF_3)-F_2(F_1^{-1}(\TCF_1)))$ (Nakas and Yiannoutsos, 2004).

The Volume under the ROC Surface (VUS) was proposed as an important index for the assessment of the diagnostic accuracy. This can be shown to be equal to $\P(X_1<X_2<X_3)$, (Mossman, 1999) and can also be calculated from the functional form of the ROC surface as $\VUS=\int_0^1 \int_0^{1-F_3(F_1^{-1}(p_1))}$ $\ROC_s(p_1, p_3) dp_3dp_1$ (Nakas and Yiannoutsos, 2004).

The ROC surface has been used in diagnostic medicine when the disease has two phases, for example, the early phase and late phase of a progressive disease. The symptoms in the early phase may be mild and ignorable, while the late phase tends to have more severe symptoms. Clearly there is an inherent ordering between healthy, early phase diseased and late phase diseased. One example is the Alzheimer's disease, which can be graded low, intermediate and high according to the progress of the disease (Chi and Zhou, 2008). 
Different medical treatments should be applied to different phases. For example, the treatment for the late phase of disease can be expensive and may even be invasive to patients, even surgeries may be included, while the treatment for the early phase can be conservative. This necessitates the identification of the two phases of the disease and thus leading to the consideration of three classes. We assume, without loss of generality, that higher test values indicate a higher level of disease.

Many methods have been proposed for estimating the ROC surface. The naive method is to replace the distribution functions in the functional form above by their empirical estimates. Kang and Tian (2013) proposed using Gaussian kernel approaches for estimation of distribution functions. A non-parametric Bayesian method of the ROC surface estimation based on Finite Polya Tree prior distributions was proposed by In\'acio et al. (2011).

A parametric method of estimating the ROC surface is based on the parametric trinormality assumption, that is, $X_1 \sim \N(\mu_1, \sigma_1^2)$, $X_2 \sim \N(\mu_0, \sigma_0^2)$, $X_3 \sim \N(\mu_2, \sigma_2^2)$. Xiong et al. (2006) obtained a closed form expression of the ROC surface under this assumption. If the data mentioned are not normally distributed, Kang and Tian (2013) proposed the use of Box-Cox to transform them to approximately normally distributed data.

Li and Zhou (2009) introduced a semi-parametric estimation of the ROC surface by generalizing the method of the estimation of the ROC curve given by Hsieh and Turnbull (1996) and Nze Ossima et al. (2013). They proposed two different estimation procedures which are based on fitting an ROC surface with the trinormal parametric form using the observed data. 

In this paper, we propose a new semiparametric method for estimating the ROC surface. This is a generalization of a Bayesian method for estimating the ROC curve using a rank-based likelihood (BRL) introduced by Gu and Ghosal (2008). We assume that, under some kind of unknown strictly monotone increasing transformation, the measurements follow three different normal distributions. Since the rank is invariant under strictly monotone increasing transformation, exploiting the rank-likelihood eliminates the need for specifying a prior distribution on the unknown transformation, and enable us to construct a Bayesian estimator of ROC surface.

We shall also consider the possibility of a verification bias in the data which widely occur in practice but is scarcely considered in the existing literature. The true disease status is verified only through the most accurate existing diagnostic test called the gold standard test. However, this kind of tests may be expensive and even invasive, so the common practice is to give this kind of verification test only to high-risk subjects which can be identified through the result of the screening test. This leads to the problem of estimating the ROC surface when not every patient's true disease status (denoted as label in this paper) is known to us. Because patients who are measured as low-risk are more likely to have their labels missing since the gold standard test will be less likely to be used for them, simply ignoring this missingness and estimating the ROC surface using only existing labels may generate biased result. To deal with the missing label, the commonly used missing at random (MAR) assumption (Little and Rubin, 1987) will be adopted. The assumption means that the probability of a subject being verified does not depend on the disease status given the observed measurements. This is reasonable in this context since the decision to obtain the gold standard test is generally made by looking at the diagnostic test result and other external factors, while the effect of the true disease status is already incorporated through its influence in diagnostic tests. 

The existing literature is limited for this problem. Chi and Zhou (2008) considered this problem for ordinal diagnostic tests and proposed the maximum likelihood estimator for the ROC surface and the VUS. To Duc et al. (2016) considered this problem for continuous diagnostic tests. They proposed several bias-corrected estimators of TCFs and thus constructed a bias-corrected ROC surface. These methods are extensions of Full Imputation (FI), Mean Score Imputation (MSI), Inverse Probability  Weighted (IPW), Semi-Parametric Efficient (SPE) estimators for the ROC curves in Alonzo and Pepe (2005). They choose suitable parametric model to compute the probability of each individual belonging to each class from verified subjects and applied this model to unverified subjects. They also use suitable parametric model to compute the probability of verification. These probabilities are used to adjust for the influence caused by missing labels.  

Through some modification, our method for estimating the ROC surface can be extended to the setting under the verification bias. This can also be regarded as a generalization of the ROC curve estimation under the verification bias proposed by Gu et al. (2014). 

The paper is organized as follows. The methodology is described in Section 2. We first consider the setting without verification bias and then consider the case of verification bias. A result describing the consistency of the posterior distribution obtained from the rank likelihood is also presented. Extensive simulation studies are conducted in Section 3. The proposed Bayesian rank likelihood method is applied to real data set in Section 4. The proof of the posterior consistency result is given in the appendix.

\section{Methodology}

\subsection{Case without verification bias}

\subsubsection{Notation}

Let us first consider the ROC surface estimation without verification bias.
Let $\bm{X}$, $\bm{Y}$ and $\bm{Z}$ denote the diagnostic measurements from the healthy, level-1 diseased and level-2 diseased groups in the study  respectively. The number of observations from healthy group, level-1 diseased group and level-2 diseased group are denoted by by $n_0$, $n_1$ and $n_2$ respectively, and the total number of subjects is $N=n_0+n_1+n_2$. 
The diagnostic measurements for all $N$ subjects in the study is denoted by $\bm{S}$=($\bm{X}$,$\bm{Y}$,$\bm{Z}$)=$\bm{S}_N$= $(S_1, ..., S_N)$ and their true disease status is denoted by $\bm{D}=(D_1,...,D_N)$.
Let $0$ mean healthy, $1$ mean level-1 diseased group and $2$ mean level-2 diseased group for $D_i$. So we have $n_0=\sum_{i=1}^N \mathbbm{1} (D_i=0)$, $n_1=\sum_{i=1}^N \mathbbm{1} (D_i=1)$, and $n_2=\sum_{i=1}^N \mathbbm{1} (D_i=2)$, where $\mathbbm{1}$ stands for the indicator function. With $\bm{D}$ and $\bm{S}$, $\bm{X, Y, Z}$ can be redefined as $\bm{X}$=$\bm{X}_{ n_0}$=$(X_{1},...,X_{n_0})$=$(S_i:D_i=0,i=1,...,N)$, $\bm{Y}$=$\bm{Y}_{n_1}$=$(Y_{1},...,Y_{n_1})$=$(S_i:D_i=1,i=1,...,N)$ and $\bm{Z}$=$\bm{Z}_{n_2}$=$(Z_{1},...,Z_{n_2})$=$(S_i:D_i=2,i=1,...,N)$. Because we assume that  there is no verification bias, the label reflects exactly the true disease status, so we have $\bm{L}=(L_1, ..., L_N)=\bm{D}$.

Since we assume that the trinomality assumption holds, under some strictly monotone increasing transformation $H$, the transformed observations will be normally distributed as described in (\ref{eq:normal}) below.
We denote the transformed measurements by $\bm{E}=H(\bm{X})$, $\bm{F}=H(\bm{Y})$, $\bm{G}=H(\bm{Z})$, and $\bm{Q}=H(\bm{S})$. Let $\widetilde{\bm{S}}$ and $\widetilde{\bm{Q}}$ stand for the order statistic of $\bm{S}$ and $\bm{Q}$ respectively. Because $H$ is a strictly monotone increasing function, we have $\bm{\widetilde{\bm{Q}}}=H(\widetilde{\bm{S}})$. In addition, let
$\widetilde{\bm{L}}$ and $\widetilde{\bm{D}}$ stand for the label and the true disease status corresponding to $\widetilde{\bm{Q}}$ respectively. The $k$th element of $\widetilde{\bm{Q}}$ and $\widetilde{\bm{S}}$ is denoted by $\widetilde{Q}_k$ and $\widetilde{S}_k$. The rank of $\bm{S}$ is defined as $\bm{R}_N=R(\bm{S})=(R(S_1), ..., R(S_N))=(R_{N1},...,R_{NN})$. Let $\Phi_{(\mu, \sigma)}(\cdot)$ denote the distribution function of $\N(\mu, \sigma^2)$ and $\phi_{(\mu, \sigma)}(\cdot)$ denote its density function. In addition, $\Phi(\cdot)=\Phi_{(0, 1)}(\cdot)$, $\phi(\cdot)=\phi_{(0, 1)}(\cdot)$.

\subsubsection{Model}
Let $F_0$, $F_1$ and $F_2$ be the continuous cumulative distribution functions of the diagnostic measurements for healthy group, level-1 disease group and level-2 disease group respectively. That is, $S_i|\{D_i=0\}\sim F_0$, $S_i|\{D_i=1\}\sim F_1$, $S_i|\{D_i=2\}\sim F_2$. Based on the trinomality assumption, under some strictly monotone increasing transformation $H$, we have:
\begin{equation}\label{eq:normal}
\begin{aligned}
Q_i|{D_i=0} &\stackrel{\mathrm{i.i.d.}}{\sim} \mathrm{N}(\mu_1, \sigma_1^2), \mu_1<0; \\
Q_i|{D_i=1} &\stackrel{\mathrm{i.i.d.}}{\sim} \mathrm{N}(0,1); \\
Q_i|{D_i=2} &\stackrel{\mathrm{i.i.d.}}{\sim} \mathrm{N}(\mu_2, \sigma_2^2), \mu_2>0;
\end{aligned}
\end{equation}
where $Q_i=H(S_i)$.

The ROC surface under trinormality is given by (Xiong et al. 2006)
\begin{equation}\label{eq:roc}
z=\Bigg[\Phi \Big(\frac{\Phi^{-1}(1-y)+d}{c} \Big)-\Phi\Big(\dfrac{\Phi^{-1}(x)+b}{a}\Big) \Bigg]_+ ,
\end{equation}
where 
\begin{equation}\label{eq:abcd}
a=1/ \sigma_1, b=\mu_1/\sigma_1, c=1/\sigma_2, d=\mu_2/\sigma_2.
\end{equation}
The volume under the ROC surface (VUS) is given by
\begin{equation}\label{eq:VUS}
\VUS=\int _{-\infty}^{\infty} \Phi(as-b)\Phi(-cs+d)\phi(s)ds.
\end{equation}

\subsubsection{Prior distribution}

In order to apply a Bayesian approach to estimate $(\mu_1, \sigma_1, \mu_2, \sigma_2)$, or equivalently, $(a,b,c,d)$, we need to specify a prior distribution for those variables first. Because it is difficult to specify a subjective prior for $(\mu_1, \sigma_1, \mu_2, \sigma_2)$, here we follow Gu et al (2014) by choosing a commonly used improper prior $\pi(\mu_1, \sigma_1, \mu_2, \sigma_2) \propto \sigma_1^{-1}\sigma_2^{-1}$ for locational-scale parameters but restricted to $\mu_1<0, \mu_2 >0$.

\subsubsection{Lemmas}

Because $H$ is a strictly monotone increasing transformation, 
$R(Q)$, the ranks of $Q$ and $L(Q)$, the labels of $Q$, preserve those of $S$. Therefore, we can define a set invariant under the action of $H$ as follows \cite{gu2009bayesian}:
\begin{equation}
\begin{aligned}\label{eq:invariant}
\mathscr{D}_{obs} =\{&(e,f,g) \in \mathbb{R}^{m+n}: R(\bm{e}, \bm{f}, \bm{g})=R(\bm{S}), L(\bm{e}, \bm{f}, \bm{g})=L(\bm{S})\},\\
=\{&(e,f,g) \in \mathbb{R}^{m+n}: \underline{e}_k <e_k <\overline{e}_k, \underline{f}_l <f_l <\overline{f}_l, \underline{g}_m <g_m <\overline{g}_m, \forall k,l,m, \\
&L(\bm{e}, \bm{f}, \bm{g})=L(\bm{S}) \},
\end{aligned}
\end{equation}
where $\bm{e}=(e_1, ..., e_{n_0}), \bm{f}=(f_1, ..., f_{n_1}), \bm{g}=(g_1, ..., g_{n_2})$, and 
\begin{equation*}
\begin{aligned}
\underline{e}_k =& \max_{i}\{e_i: R_{N_i} < R_{N_k}\} \vee \max_j\{f_j: R_{N_{n_0+j}} < R_{N_k}\} \vee \max_h\{g_h: R_{N_{n_0+n_1+h}} < R_{N_k}\}  \\
\overline{e}_k =& \min_{i}\{e_i: R_{N_i} > R_{N_k}\} \wedge \min_j\{f_j: R_{N_{n_0+j}} > R_{N_k}\} \wedge \min_h\{g_h: R_{N_{n_0+n_1+h}} > R_{N_k}\}  \\
\underline{f}_l =& \max_{i}\{e_i: R_{N_i} < R_{N_{n_0+l}}\} \vee \max_j\{f_j: R_{N_{n_0+j}} < R_{N_{n_0+l}}\} \\
&\vee \max_h\{g_h: R_{N_{n_0+n_1+h}} < R_{N_{n_0+l}}\}  \\
\overline{f}_l =& \min_{i}\{e_i: R_{N_i} > R_{N_{n_0+l}}\} \wedge \min_j\{f_j: R_{N_{n_0+j}} > R_{N_{n_0+l}}\} \\
&\wedge \min_h\{g_h: R_{N_{n_0+n_1+h}} > R_{N_{n_0+l}}\} \\
\underline{g}_m =& \max_{i}\{e_i: R_{N_i} < R_{N_{n_0+n_1+m}}\} \vee \max_j\{f_j: R_{N_{n_0+j}} < R_{N_{n_0+n_1+m}}\}  \\
&\vee  \max_h\{g_h: R_{N_{n_0+n_1+h}} < R_{N_{n_0+n_1+m}}\}  \\
\overline{g}_m =& \min_{i}\{e_i: R_{N_i} > R_{N_{n_0+n_1+m}}\} \wedge \min_j\{f_j: R_{N_{n_0+j}} > R_{N_{n_0+n_1+m}}\} \\
&\wedge \min_h\{g_h: R_{N_{n_0+n_1+h}} > R_{N_{n_0+n_1+m}}\} \\
\end{aligned}
\end{equation*}
for all $k, i =1,...,n_0$; $l, j=1,...,n_1$, $m, h=1,...,n_2$. In the above, for am empty index set the maximum is interpreted as $-\infty$ and the minimum as $\infty$.

\subsubsection{Likelihood and posterior distribution}

First we consider the case without verification bias. Based on invariant set (\ref{eq:invariant}), a likelihood is given by
\begin{equation}
\begin{aligned}
& \P \{(\bm{E, F, G}) \in \mathscr{D}_{obs} | \mu_1, \sigma_1, \mu_2, \sigma_2\} \\
= & \P \{(\bm{e,f,g}) \in \mathbb{R}^{n_0+n_1+n_2}: R(\bm{e,f,g}) = R(\bm{S}), L(\bm{e,f,g})=L(\bm{S}) | \mu_1, \sigma_1, \mu_2, \sigma_2 \}.
\end{aligned}
\end{equation}

Notice that without verification bias, $\widetilde{\bm{L}}=\widetilde{\bm{D}}$, i.e., the labels reflect the true disease status. Using Bayesian approach, the posterior sampling can be done by Gibbs sampling as follows:

\begin{itemize}
\item Choose an initial value of $(\mu_1, \sigma_1, \mu_2, \sigma_2)$. Generate initial value for $(Q_1, ..., Q_N)$ according to (\ref{eq:normal}), i.e., a sample which contains $n_0$ many random variables from $\N(\mu_1, \sigma_1^2)$, $n_1$ many from $\N(0,1)$ and $n_2$ many from $\N(\mu_2, \sigma_2^2)$, where $n_0$, $n_1$ and $n_2$ are the number of observations labeled as healthy, level-1 and level-2 diseased groups.

\item Iteratively execute the following steps:

\begin{enumerate}
\item Sample $\widetilde{\bm{Q}}$ sequentially conditional on $(\mu_1, \sigma_1, \mu_2, \sigma_2)$ by
\begin{equation} \label{eq:Q}
\widetilde{Q}_i|\mbox{rest} \sim 
\begin{cases}
\mathrm{TN}(\mu_1, \sigma_1^2, (\widetilde{Q}_{i-1}, \widetilde{Q}_{i+1})), &\mbox{ if } \widetilde{D}_{i}=0, \\
\mathrm{TN}(0, 1, (\widetilde{Q}_{i-1} , \widetilde{Q}_{i+1})), &\mbox{ if } \widetilde{D}_{i}=1, \\
\mathrm{TN}(\mu_2, \sigma_2^2, (\widetilde{Q}_{i-1} , \widetilde{Q}_{i+1})), &\mbox{ if } \widetilde{D}_{i}=2, \\
\end{cases}
\end{equation}
where $i=1,...,N$, $\widetilde{Q}_0=-\infty$, $\widetilde{Q}_{N+1}=\infty$.

\item Sample $(\mu_1, \sigma_1, \mu_2, \sigma_2)$ conditional on $\widetilde{\bm{Q}}$ and $\widetilde{\bm{D}}$  by
\begin{equation}\label{eq:musigma}
\begin{aligned}
\mu_1|\mbox{rest} &\sim \mathrm{TN}(\bar{E}_{n_0}, \sigma_1^2/n_0, (-\infty, 0)), \\
\sigma_1^2|\mbox{rest} &\sim \mathrm{IG}((n_0-1)/2, (n_0-1)s_0^2/2), \\
\mu_2|\mbox{rest} &\sim \mathrm{TN}(\bar{G}_{n_2}, \sigma_2^2/n_2, (0, \infty)),\\
\sigma_2^2|\mbox{rest} &\sim \mathrm{IG}((n_2-1)/2, (n_2-1)s_2^2/2), \\
\end{aligned}
\end{equation}
where $\mathrm{IG}$ stands for inverse gamma distribution, $\mathrm{TN}$ stands for truncated normal distribution, $s_0^2=\sum_{j=1}^{n_0}(E_j-\bar{E}_{n_0})^2/(n_0-1)$, $\bar{E}_{n_0}=\sum_{j=1}^{n_0}E_j/n_0$;  $s_2^2=\sum_{j=1}^{n_2}(G_j-\bar{G}_{n_2})^2/(n_2-1)$ and $\bar{G}_{n_2}=\sum_{j=1}^{n_2}G_j/n_2$.

\item Calculate $a, b, c, d$ according to (\ref{eq:abcd}) and VUS according to (\ref{eq:VUS}).
\end{enumerate}

\item Monitor the convergence through trace plot and discard all samples for a suitable burn-in period, we obtain the estimates of the parameters  $(\hat{a}, \hat{b}, \hat{c}, \hat{d})$ in the parametric ROC surface by averaging out the sample values of $(a,b,c,d)$ in each iteration respectively. The VUS is also estimated by averaging out the computed value of the VUS in each MCMC iteration. 

\end{itemize}

\subsection{Case with verification bias}

\subsubsection{Notation}

The notation is mostly the same as defined in Section 2.1.1, the only difference is with the label.
Under verification bias, only a fraction of patients have their true disease status $D_i$ observed, $i=1,...,N$, so $L=(L_1, ..., L_N)$ is different from the previous case. With verification bias, $L_i$ is defined by 
\begin{equation}\label{eq:label}
L_i=
\begin{cases}
0, \mbox{if label is observed and } D_i=0, \\
1, \mbox{if label is observed and } D_i=1, \\
2, \mbox{if label is observed and } D_i=2, \\
3, \mbox{if label is not observed.}
\end{cases}
\end{equation}
Thus $L_i$ indicates missing status as well as true disease status if the label is actually observed. 

Also we need to define some new variables to adjust for the verification bias. Let $n_0^*$, $n_1^*$ and $n_2^*$ stand for the number of observations labeled healthy, level-1 and level-2 diseased groups respectively, i.e., $n_0^*=\sum_{i=1}^N \mathbbm{1} (L_i=0)$, $n_1^*=\sum_{i=1}^N \mathbbm{1} (L_i=1)$, and $n_2^*=\sum_{i=1}^N \mathbbm{1} (L_i=2)$, and put $N^*=n_0^*+n_1^*+n_2^*$, the total number of observations having labels. Note that $\bm{X}$, $\bm{Y}$ and $\bm{Z}$ are not observable in this case since $\bm{D}$ is not observable. In addition, we define the collection of unobserved labels as $\bm{D}_{\mathrm{un}}=\{D_i:L_i=3,i\leq N\}$ and the collection of observed labels as $\bm{D}_{\mathrm{obs}}=\{D_i:L_i=0,1 \ \mbox{or} \ 2,i\leq N\}$.

\subsubsection{Model}
Assume that, the disease prevalence rates for level-1 and level-2 in the population are $\lambda_1$ and $\lambda_2$ respectively. So the true number of subjects in each group follows $(n_0, n_1, n_2) \sim \Mult(N, (\lambda_0, \lambda_1, \lambda_2))$, where $\Mult$ stands for the Multinomial distribution, $\lambda_0=1-\lambda_1-\lambda_2$, $0<\lambda_1<1$, $0<\lambda_2<1$, and $0<\lambda_1+\lambda_2<1$. 

Because of the existence of the verification bias, we need to build a model for observing actual labels.
In a clinical practice, gold standard test will typically be prescribed according to the screening test results. 
To be specific, a subject with higher risk of disease, suggested by higher score in the diagnostic test, is more likely to be forwarded to the gold standard test which is more thorough and more accurate. 
Because of this, missing completely at random may not be appropriate.
Simply ignoring unverified subjects will create bias to our estimation. Here we follow Gu et al. (2014) and model this as missing at random. 
In general, this model of verifying disease status can be represented as
\begin{equation}\label{eq:vb}
\P(L_i \neq 3|Q_i, D_i)=g(Q_i),
\end{equation}
for a given monotone increasing function $g$. Note that $Q_i$s here are not observed since we can only observe $S_i=H(Q_i)$ and $H$ is unknown.

In this paper, we consider two reasonable missing mechanisms following Gu and Ghoshal (2014).  The first model is proposed by Alonzo and Pepe (2005):
\begin{equation}\label{eq:threshold}
\P(L_i \neq 3|S)=
\begin{cases}
1, \ \ & \mbox{if } S>S_{(p_1N)} ,\\
p_2, \ \ & \mbox{otherwise}.
\end{cases}
\end{equation}
Here, $p_1$ and $p_2$ are known probabilities. Since  the ordering will not change after transformation $H$, $Q$'s and $S$'s share the same ordering, so (\ref{eq:threshold}) can be considered as a special case of (\ref{eq:vb}).
\begin{equation*}
g(Q)=
\begin{cases}
1, \ \ &\mbox{if }  Q>Q_{(p_1N)}, \\ 
p_2, \ \ &\mbox{if } Q\leq Q_{(p_1N)}.  
\end{cases}
\end{equation*}
Another model exploits the probit link:
\begin{equation}\label{eq:probit}
\P(L_i \neq 3|Q_i)=\Phi(\alpha+\beta Q_i)
\end{equation}
where $\alpha$, $\beta$ are known parameters with $\beta>0$.

Notice that the case without the verification bias can be regarded as a special case of (\ref{eq:vb}) by simply setting $g(Q_i)=1$. 

\subsubsection{Prior distribution}

We will apply a Bayesian approach to estimate $(\lambda_1, \lambda_2, \mu_1, \sigma_1, \mu_2, \sigma_2)$, or equivalently,  $(\lambda_1, \lambda_2,  a,b,c,d)$. We still use the improper prior $\pi(\mu_1, \sigma_1, \mu_2, \sigma_2) \propto \sigma_1^{-1}\sigma_2^{-1}$ for $(\mu_1, \sigma_1, \mu_2, \sigma_2)$. For disease prevalence rates, the prior is set to be $(\lambda_0, \lambda_1, \lambda_2) \sim \mathrm{Dir}(\alpha_0, \alpha_1, \alpha_2)$, where $\Dir(\cdot)$ stands for the Dirichlet distribution, $\alpha_0$, $\alpha_1$ and $\alpha_2$ are chosen according to the mean and standard error from our prior knowledge.

\subsubsection{Lemmas} 
The following lemmas are used to obtain posterior distribution and to prove the posterior consistency.
The proofs are deferred to the Appendix.

\begin{lemma} \label{lemma2}
Let $D_1, ..., D_N$ be the true disease status corresponding to $N$ subjects with independent diagnostic measurements $S_1, ..., S_N$. Under the verification bias, assume that $(n_0, n_1, n_2) \sim \Mult(N, (\lambda_0, \lambda_1, \lambda_2))$, and that (\ref{eq:normal}) and (\ref{eq:vb}) hold. Then we have
\begin{equation} \label{eq:Ddist}
\begin{aligned}
\P(D_i=0| Q_i=t, L_i=3) &= \P(D_i=0| Q_i=t, L_i \neq 3) \\
&=\dfrac{\lambda_0 \phi_{(\mu_1, \sigma_1)}(t)}{\lambda_0 \phi_{(\mu_1, \sigma_1)}(t )+\lambda_1 \phi(t )+\lambda_2 \phi_{(\mu_2, \sigma_2)}(t )}, \\
\P(D_i=1| Q_i=t, L_i=3) &= \P(D_i=1| Q_i=t, L_i \neq 3) \\
&=\dfrac{\lambda_1 \phi(t)}{\lambda_0 \phi_{(\mu_1, \sigma_1)}(t )+\lambda_1 \phi(t)+\lambda_2 \phi_{(\mu_2, \sigma_2)}(t )},\\
\P(D_i=2| Q_i=t, L_i=3) &= \P(D_i=2| Q_i=t, L_i \neq 3) \\ 
&=\dfrac{\lambda_2 \phi_{(\mu_2, \sigma_2)}(t)}{\lambda_0 \phi_{(\mu_1, \sigma_1)}(t)+\lambda_1 \phi(t)+\lambda_2 \phi_{(\mu_2, \sigma_2)}(t)}.
\end{aligned}
\end{equation}
where $L_i$ is defined in (\ref{eq:label}).
\end{lemma}

\begin{remark}
For the case where there is no verification bias, i.e., $L_i \neq 3$, then instead of (\ref{eq:Ddist}) we have
\begin{equation}
\begin{aligned}
\P(D_i=0| Q_i=t) &=\dfrac{\lambda_0 \phi_{(\mu_1, \sigma_1)}(t)}{\lambda_0 \phi_{(\mu_1, \sigma_1)}(t )+\lambda_1 \phi(t )+\lambda_2 \phi_{(\mu_2, \sigma_2)}(t )}, \\
\P(D_i=1| Q_i=t) &=\dfrac{\lambda_1 \phi(t)}{\lambda_0 \phi_{(\mu_1, \sigma_1)}(t )+\lambda_1 \phi(t)+\lambda_2 \phi_{(\mu_2, \sigma_2)}(t )},\\
\P(D_i=2| Q_i=t) &=\dfrac{\lambda_2 \phi_{(\mu_2, \sigma_2)}(t)}{\lambda_0 \phi_{(\mu_1, \sigma_1)}(t)+\lambda_1 \phi(t)+\lambda_2 \phi_{(\mu_2, \sigma_2)}(t)}.
\end{aligned}
\end{equation}
\end{remark}

Under verification bias, Lemma \ref{lemma2} shows that given $Q$, the random variable $D$ and the occurrence of $\{L_i \neq 3\}$ are independent. In addition, we can see that  the expressions above are free of $g(\cdot)$, which is the model of verification. Thus to calculate the posterior distribution we do not actually need to know $g$.
The monotonicity of $g$ is needed because $Q$ is unobservable. Given monotonicity of $g$ allows (\ref{eq:vb}) to be presented in form of $S$. 

Thus, this method is guarded against misspecification of the verification model. Thus, our method is robust compared with other methods which focus on verification probability.

The following lemma has nothing to do with computing the posterior, but will  be used to study the consistency of posterior for large sample.

\begin{lemma} \label{lemma3}
Under verification bias, assume that (\ref{eq:normal}) and (\ref{eq:vb}) hold. Then the conditional density of $Q$ given that it is verified is given by
\begin{equation}
\begin{aligned}
f_Q(t|L \neq 3)= & (1-\lambda_{1(\mu_1, \sigma_1, \mu_2, \sigma_2, g)}^*-\lambda_{2(\mu_1, \sigma_1, \mu_2, \sigma_2, g)}^*)\phi^*_{(\mu_1, \sigma_1, g)}(t) + \lambda_{1(\mu_1, \sigma_1, \mu_2, \sigma_2, g)}^* \phi_{(g)}^*(t) \\
& + \lambda_{2(\mu_1, \sigma_1, \mu_2, \sigma_2, g)}^*\phi^*_{(\mu_1, \sigma_1, g)}(t),
\end{aligned}
\end{equation} 
where $\lambda_{1(\mu_1, \sigma_1, \mu_2, \sigma_2, g)}^*$, $\lambda_{2(\mu_1, \sigma_1, \mu_2, \sigma_2, g)}^*$ and $\phi^*_{(\mu, \sigma, g)}(t)$ are defined by
\begin{equation}
\begin{aligned}\label{def:lambda}
\lambda_{1(\mu_1, \sigma_1, \mu_2, \sigma_2, g)}^* & = \dfrac{\lambda_1 \int g(s)\phi(s) ds}{\int g(s)\{(1-\lambda_1-\lambda_2)\phi_{(\mu_1, \sigma_1)}(s)+\lambda_1\phi(s)+\lambda_2 \phi_{(\mu_2, \sigma_2)}(s)\} ds}, \\ 
\lambda_{2(\mu_1, \sigma_1, \mu_2, \sigma_2, g)}^* & = \dfrac{\lambda_2 \int g(s)\phi_{(\mu_2, \sigma_2)}(s) ds}{\int g(s)\{(1-\lambda_1-\lambda_2)\phi_{(\mu_1, \sigma_1)}(s)+\lambda_1\phi(s)+\lambda_2 \phi_{(\mu_2, \sigma_2)}(s)\} ds}, \\ 
\phi^*_{(\mu, \sigma, g)}(t) & = \dfrac{g(t) \phi_{(\mu, \sigma)}(t)}{\int g(s) \phi_{(\mu, \sigma)}(s) ds}, 
\end{aligned}
\end{equation} 
and $\phi_{(g)}^*(t)$ stands for $\phi_{(0,1,g)}^*(t)$.
\end{lemma}

\subsubsection{Posterior distribution}

Our method is still based on the rank-based partial likelihood on the invariant set (\ref{eq:invariant}). However, we need to do some adjustment since now $\widetilde{\bm{D}}$ is no longer equal to $\widetilde{\bm{L}}$ and is only partially known. More specifically, we have $D_i=L_i$ where $D_i \in \bm{D}_{\mathrm{obs}}$, but $D_i \in \bm{D}_{\mathrm{un}}$ are not observable and are unknown to us. Thus we need to apply data augmentation technique to get $\widetilde{\bm{Q}}$ and $\bm{D}_{\mathrm{un}}$, and then implement Gibbs sampling to compute the posterior distribution conditional on the remaining variables. The computing algorithm is as follows:

\begin{itemize}
\item Initialize missing labels $D_i \in \bm{D}_{\mathrm{un}}$ using $(N-N^*)$ independent $\Mult(1, (\lambda_0, \lambda_1, \lambda_2))$ variables, where $(\lambda_0, \lambda_1, \lambda_2)$ is generated from $\Dir(\alpha_0, \alpha_1, \alpha_2)$ to form an initial complete $\widetilde{\bm{D}}$. The number of subjects in each class can be calculated as $n_0'=\Sigma_{i=1}^N \mathbbm{1}\{\widetilde{\bm{D}}_i=0\}$, $n_1'=\Sigma_{i=1}^N \mathbb m{1}\{\widetilde{\bm{D}}_i=1\}$ and $n_2'=\Sigma_{i=1}^N \mathbbm{1}\{\widetilde{\bm{D}}_i=2\}$

\item Choose an initial value of $(\mu_1, \sigma_1, \mu_2, \sigma_2)$. Generate an initial value for $(Q_1, ..., Q_N)$ according to (\ref{eq:normal}), i.e., a sample which contains $n_0'$ many random variables from $\N(\mu_1, \sigma_1^2)$, $n_1'$ many from $\N(0,1)$ and $n_2'$ many from $\N(\mu_2, \sigma_2^2)$.

\item Iteratively execute the following steps:

\begin{enumerate}
\item Sample $\widetilde{\bm{Q}}$ sequentially conditional on $(\mu_1, \sigma_1, \mu_2, \sigma_2)$ by
\begin{equation} \label{eq:Q2}
\widetilde{Q}_i |\mbox{rest} \sim 
\begin{cases}
\TN(\mu_1, \sigma_1^2, (\widetilde{Q}_{i-1} , \widetilde{Q}_{i+1})), &\mbox{ if } \widetilde{D}_{i}=0, \\
\TN(0, 1, (\widetilde{Q}_{i-1} , \widetilde{Q}_{i+1})), &\mbox{ if }  \widetilde{D}_{i}=1, \\
\TN(\mu_2, \sigma_2^2, (\widetilde{Q}_{i-1} , \widetilde{Q}_{i+1})), &\mbox{ if }  \widetilde{D}_{i}=2, \\
\end{cases}
\end{equation}
where $i=1,...,N$, $\widetilde{Q}_0=-\infty$, $\widetilde{Q}_{N+1}=\infty$. 

\item Sample $(\mu_1, \sigma_1, \mu_2, \sigma_2)$ conditional on $\widetilde{\bm{Q}}$ and $\widetilde{\bm{D}}$  by
\begin{equation} \label{eq:musigma2}
\begin{aligned}
\mu_1|\mbox{rest}  &\sim \mathrm{TN}(\bar{E}_{n_0'}, \sigma_1^2/n_0', (-\infty, 0)), \\
\sigma_1^2|\mbox{rest} &\sim \mathrm{IG}((n_0'-1)/2, (n_0'-1)s_0^2/2), \\
\mu_2|\mbox{rest}  &\sim \mathrm{TN}(\bar{G}_{n_2'}, \sigma_2^2/n_2', (0, \infty)),\\ 
\sigma_2^2|\mbox{rest}  &\sim \mathrm{IG}((n_2'-1)/2, (n_2'-1)s_2^2/2), 
\end{aligned}
\end{equation}
where $n_0'=\sum_{i=1}^N\{ \mathbbm{1}(\tilde{D_i}=0, \tilde{L_i}=3)+ \mathbbm{1}(\tilde{L}_i=0)\}$, $s_0^2=\sum_{j=1}^{n_0'}(E_j-\bar{E}_{n_0'})^2/(n_0'-1)$, $\bar{E}_{n_0'}=\sum_{j=1}^{n_0'}E_j/n_0'$; $n_2'=\sum_{i=1}^N\{ \mathbbm{1}(\tilde{D_i}=2, \tilde{L_i}=3)+ \mathbbm{1}(\tilde{L}_i=2)\}$, $s_2^2=\sum_{j=1}^{n_2'}(G_j-\bar{G}_{n_2'})^2/(n_2'-1)$ and $\bar{G}_{n_2'}=\sum_{j=1}^{n_2'}G_j/n_2'$.

\item Sample $(\lambda_0, \lambda_1, \lambda_2)$ given the rest by
\begin{equation} \label{eq:postlambda}
(\lambda_0, \lambda_1, \lambda_2)|\mbox{rest} \sim \Dir(\alpha_0+n_0', \alpha_1+n_1', \alpha_2+n_2'),
\end{equation}
where $n_1'=\sum_{i=1}^N\{ \mathbbm{1}(\tilde{D_i}=1, \tilde{L_i}=3)+ \mathbbm{1}(\tilde{L}_i=1)\}$.

\item Sample the augmentation variable $\tilde{D}_i \in \bm{D}_{\mathrm{un}}$ by
\begin{equation} \label{eq:postD}
\begin{aligned}
\Big(\mathbbm{1}\{\tilde{D}_i =0\}, \mathbbm{1}\{\tilde{D}_i =1\}, \mathbbm{1}\{\tilde{D}_i =2\} \Big)|\mbox{rest} \sim \\
\Mult \Bigg(1, \bigg(\dfrac{\lambda_0 \phi_{(\mu_1, \sigma_1)}(\widetilde{Q}_i  )}{\lambda_0 \phi_{(\mu_1, \sigma_1)}(\widetilde{Q}_i  )+\lambda_1 \phi(\widetilde{Q}_i  )+\lambda_2 \phi_{(\mu_2, \sigma_2)}(\widetilde{Q}_i  )}, \\
\dfrac{\lambda_1 \phi(\widetilde{Q}_i  )}{\lambda_0 \phi_{(\mu_1, \sigma_1)}(\widetilde{Q}_i  )+\lambda_1 \phi(\widetilde{Q}_i  )+\lambda_2 \phi_{(\mu_1, \sigma_1)}(\widetilde{Q}_i  )}, \\
\dfrac{\lambda_2 \phi_{(\mu_2, \sigma_2)}(\widetilde{Q}_i  )}{\lambda_0 \phi_{(\mu_1, \sigma_1)}(\widetilde{Q}_i  )+\lambda_1 \phi(\widetilde{Q}_i  )+\lambda_2 \phi_{(\mu_2, \sigma_2)}(\widetilde{Q}_i  )} \bigg) \Bigg)
\end{aligned}
\end{equation}

\item Calculate $a, b, c, d$ according to (\ref{eq:abcd}) and VUS according to (\ref{eq:VUS}).
\end{enumerate}

\item Monitor the convergence through trace plot and discard all samples for a suitable burn-in period, we obtain the estimates of the parameters  $(\hat{a}, \hat{b}, \hat{c}, \hat{d})$ in the parametric ROC surface by averaging out the sample values of $(a,b,c,d)$ in each iteration respectively. The VUS is also estimated by averaging out the computed value of the VUS in each MCMC iteration. 

\end{itemize}

\subsection{Consistency of the posterior distributions}

As we already showed that the case without verification bias can be regarded as a special case of under verification bias model by simply setting $g(Q_i)=1$ in (\ref{eq:vb}). Thus we only need to show posterior consistency under verification bias.

Let $\nu$ denote the Lebesgue norm and $\pi$ denote the prior density for $(\mu_1, \sigma_1, \mu_2, \sigma_2)$ with respect to $\nu$.
Let $(\mu_{1}^*, \sigma_{1}*, \mu_{2}^*, \sigma_{2}^*)$ be the true value of $(\mu_1, \sigma_1, \mu_2, \sigma_2)$. We shall show that the posterior distribution $\Pi(\cdot|R(\bm{S}), \bm{L})$ of $(\mu_1, \sigma_1, \mu_2, \sigma_2)$ generally concentrates near $(\mu_{1}^*, \sigma_{1}^*, \mu_{2}^*, \sigma_{2}^*)$ for large sample size. 

\begin{theorem} \label{consistency}
Assume (\ref{eq:normal}) and (\ref{eq:vb}) hold for a monotone increasing function $g$: $\mathbb{R}$ to $(0,1)$. Let $\pi(\mu_1, \sigma_1, \mu_2, \sigma_2) >0$ a.e. $[\nu]$. $ \mathbb{R}^- \times \mathbb{R}^+ \times \mathbb{R}^+ \times \mathbb{R}^+$. Then for $(\mu_{1}^*, \sigma_{1}^*, \mu_{2}^*, \sigma_{2}^*) \ a.e. \ [\nu]$, and any neighborhood $\mathscr{U}_0$ of  $(\mu_{1}^*, \sigma_{1}*, \mu_{2}^*, \sigma_{2}^*)$ , we have that
\begin{equation}
\lim_{N \rightarrow \infty} \Pi((\mu_1, \sigma_1, \mu_2, \sigma_2)\in \mathscr{U}_0 | R(\bm{S}), \bm{L})=1, a.s.
\end{equation}
with respect to  $[P_{\mu_{1}^*, \sigma_{1}*, \mu_{2}^*, \sigma_{2}^*, g, H}^{\infty}]$, the true joint distribution of all ranks and labels.
\end{theorem}

We prove that this theorem using a general posterior consistency theorem by Doob.  The complete proof is given in appendix. The main idea is that we show that the parameters can be expressed as a function of the  observations and then apply Doob's theorem to show posterior consistency. The theorem provides a theoretical justification of our method from the frequentist point of view.

\section{Simulation}

\subsection{Without verification bias}

In this simulation, set $n_0=n_1=n_2=n$, so we generate $N=3n$ data in total with $n$ data from each category as the unobserved after-transformation measurements, where $n$ is set to be 50 and 100. and the true underlying normal distribution is taken to be $\N(-1.8, 1.5)$, $\N(0,1)$ and $\N(2,2)$, i.e., the true value for $(a,b,c,d)$ in (\ref{eq:roc}) is $(0.667, -1.2, 0.5, 1)$ and the true VUS=0.671. Also we consider two different true transformations $H$ here, the first is logarithm  transformation, i.e.,  $H(x)=\log(x)$; the second is logit transformation, i.e., $H(x)=\log({x}/(1-x))$. We first simulate unobserved after-transformation measurements, and then use the inverse of the true transformation to get the raw data that we observe in reality. 

For comparison purpose, we compare our methods (BRL) to existing state-of-art methods. The first method we consider is based on eliminating the unknown transformation $H$. Here, we follow Kang and Tian (2013) to use the Box-Cox transformation. Notice that one of the first true transformations we choose here, i.e., the logarithmic transformation, belongs to the Box-Cox transformation family. After Box-Cox transformation, we use the maximum likelihood method to estimate parameters following the idea of Xiong et al. (2006).

Also we compare BRL to two other semiparametric methods (denoted here by Semi1 and Semi2) proposed by Li and Zhou (2009). The simulation results are shown in Tables 1 and 2. We calculate the bias and the MSE of the estimates of the parameters and also the bias and the MSE of the estimates of the volume under the surface (VUS). The bias and the MSE shown in the table are their true values multiplied by $10^2$ and the Monte Carlo standard error for each estimate is shown in parentheses below (also multiplied by $10^2$).

All the estimates are based on $100$ simulated data sets. For each data set, our BRL method is calculated with $90000$ Gibbs samples ($100000$ MCMC iterations after $10000$ samples used for burn-in). 

\begin{table}[]
\centering
\caption{Bias $(\times 10^{2})$ and MSE $(\times 10^{2})$ for estimates of parameter values without verification bias}
\label{my-label}
\begin{tabular}{rrrrrrrrrr}
\multicolumn{2}{l}{\multirow{3}{*}{}}  & \multicolumn{4}{l}{$n=50$}     & \multicolumn{4}{l}{$n=100$}   \\ \cmidrule(lr){3-6}\cmidrule(lr){7-10}
\multicolumn{2}{l}{}                  & BRL    & Box-Cox   & Semi1   & Semi2   & BRL     & Box-Cox   & Semi1   & Semi2  \\ \hline
\multicolumn{2}{l}{}                  & \multicolumn{8}{l}{true transformation=$\log(x)$} \\ \hline
\multirow{4}{*}{$a$}       & \multirow{2}{*}{bias}      
&  $-19.5$ &$6.0$ &$-15.3$ &$-12.2$ &$-10.7$ &$4.6$ &$-16.1$ &$-12.9$\\
&& $(1.6)$ &$(1.3)$&$(1.3)$ &$(1.4)$ & $(0.8)$ & $(0.9)$ & $(0.9)$ &$(1.4)$ \\
                         & \multirow{2}{*}{MSE}       
& $6.4$  &$2.0$ &$4.0$ &$3.4$ &$1.3$ &$1.1$ &$3.5$ &$3.7$ \\
&& $(0.7)$ & $(0.4)$ &$(0.4)$ &$(0.3)$ & $(0.2)$ &$(0.2)$ & $(0.3)$ &$(0.3)$\\
\multirow{4}{*}{$b$}       & \multirow{2}{*}{bias}       
& $4.0$ &$-0.1$ &$26.6$ &$27.1$ &$1.3$ &$-4.6$ &$19.2$ &$21.7$\\
&& $(2.2)$ & $(2.4)$ &$(2.0)$ &$(1.7)$ & $(1.3)$ &$(1.7)$ & $(1.4)$ &$(1.2)$\\
                         & \multirow{2}{*}{MSE}        
& $4.8$ &$5.6$ &$11.0$ &$10.2$ &$1.7$ &$2.9$ &$5.8$ &$6.3$ \\
&& $(0.8)$ & $(0.8)$ &$(1.2)$ &$(1.1)$ &$(0.2)$ &$(0.5)$ &  $(0.5)$ &$(0.5)$\\
\multirow{4}{*}{$c$}       & \multirow{2}{*}{bias}       
& $-14.4$ &$-3.6$ &$17.4$ &$17.0$ &$-6.0$ &$-2.8$ &$21.4$ &$20.6$ \\
&& $(1.2)$ & $(1.0)$ &$(1.3)$ &$(1.3)$ &$(0.7)$ &$(0.7)$ & $(1.5)$ & $(1.4)$\\
                         & \multirow{2}{*}{MSE}        
& $3.6$ &$1.2$ &$4.7$ &$4.6$ &$0.9$ &$0.6$ &$6.9$ &$6.1$ \\
&& $(0.4)$ & $(0.2)$ &$(0.5)$ &$(0.6)$ &$(0.1)$ & $(0.1)$ &  $(0.9)$ & $(0.8)$\\
\multirow{4}{*}{$d$}       & \multirow{2}{*}{bias}       
& $-3.3$ &$-2.8$ &$28.9$ &$24.1$ &$-1.2$ &$-0.2$ &$27.8$ &$24.5$ \\
&& $(1.8)$ & $(1.9)$ &$(2.0)$ &$(1.9)$ &$(1.4)$ & $(1.4)$ &  $(1.9)$ & $(1.7)$ \\
                         & \multirow{2}{*}{MSE}        
& $3.3$ &$3.6$ &$12.5$ &$9.4$ &$1.9$ &$2.1$ &$11.3$ &$8.9$\\
&& $(0.4)$& $(0.5)$ &$(1.4)$ &$(1.2)$ &$(0.3)$ &$(0.4)$ & $(1.3)$ & $(1.1)$\\
\hline
\multicolumn{2}{l}{}                  & \multicolumn{8}{l}{true transformation=$\log({x}/({1-x}))$} \\ \hline
\multirow{4}{*}{$a$}       & \multirow{2}{*}{bias}       
&$-16.0$ &$25.5$ &$-15.6$ &$-12.1$ &$-8.7$ &$28.2$ &$-18.1$ &$-13.5$\\
&& $(1.4)$& $(1.7)$ &$(1.3)$ &$(1.7)$ &$(1.0)$ & $(1.3)$ & $(0.8)$ &$(1.3)$ \\
                         & \multirow{2}{*}{MSE}        
&$4.4$ &$9.3$ &$4.1$ &$4.2$ &$1.8$ &$9.7$ &$4.0$ &$3.6$ \\
&&$(0.5)$ & $(1.0)$ &$(0.5)$ &$(0.4)$ & $(0.3)$ &$(8.0)$ & $(0.3)$ &$(0.3)$ \\
\multirow{4}{*}{$b$}       & \multirow{2}{*}{bias}       
&$0.3$ &$-14.1$ &$23.5$ &$26.3$ &$0.8$ &$-22.3$ &$22.1$ &$23.7$\\
&& $(2.1)$& $(3.1)$ &$(2.1)$ &$(1.7)$ & $(1.6)$&$(2.5)$ & $(1.4)$ &$(1.2)$ \\
                         & \multirow{2}{*}{MSE}        
&$4.3$ &$11.6$ &$9.9$ &$9.7$ &$2.6$ &$10.9$ &$6.8$ &$7.2$ \\
&&$(0.6)$ & $(2.0)$ &$(1.1)$ &$(1.1)$ & $(0.6)$&$(1.7)$ & $(0.7)$ &$(0.6)$ \\
\multirow{4}{*}{$c$}       & \multirow{2}{*}{bias}       
&$-10.9$ &$38.1$ &$18.6$ &$17.2$ &$-5.1$ &$36.9$ &$19.5$ &$19.4$\\
&&$(1.1)$ & $(1.5)$ &$(1.3)$ &$(1.4)$ &$(0.9)$ &$(1.1)$ & $(1.3)$ &$(1.2)$ \\
                         & \multirow{2}{*}{MSE}        
&$2.4$ &$16.7$ &$5.2$ &$5.1$ &$1.0$ &$14.8$ &$5.7$ &$5.3$ \\
&&$(0.3)$ & $(1.4)$ &$(0.6)$ &$(0.7)$ &$(0.1)$ &$(0.9)$ & $(0.8)$ &$(0.6)$ \\
\multirow{4}{*}{$d$}       & \multirow{2}{*}{bias}       
&$-1.6$ &$9.0$ &$28.9$ &$22.5$ &$1.3$ &$10.5$ &$27.0$ &$24.1$ \\
&&$(2.0)$ & $(3.4)$ &$(1.8)$ &$(1.7)$ &$(1.5)$ &$(2.5)$ & $(1.8)$ &$(1.6)$ \\
                         & \multirow{2}{*}{MSE}        
&$4.1$ &$12.1$ &$11.7$ &$7.8$ &$2.3$ &$7.1$ &$10.7$ &$8.5$\\
&&$(0.5)$ & $(3.0)$ &$(1.1)$ &$(0.8)$ &$(0.3)$ &$(1.5)$ & $(1.2)$ &$(1.0)$ \\
\hline
\end{tabular}
\end{table}

\begin{table}[]
\centering
\caption{Bias $(\times 10^{2})$ and MSE $(\times 10^{2})$ for estimates of VUS without verification bias}
\label{my-label}
\begin{tabular}{rrrrrrrrrr}
\multicolumn{2}{l}{\multirow{3}{*}{}}  & BRL  & Box-Cox   & Semi1   & Semi2   & BRL  & Box-Cox   & Semi1   & Semi2   \\ \cmidrule(lr){3-6} \cmidrule(lr){7-10}
\multicolumn{2}{l}{}             & \multicolumn{4}{l}{true transformation=$\log(x)$}        \\ \hline
\multicolumn{2}{l}{}            & \multicolumn{4}{l}{$n=50$}  & \multicolumn{4}{l}{$n=100$}   \\  \hline
\multirow{4}{*}{VUS}       &\multirow{2}{*}{bias}       
&$1.0$  &$-1.6$ &$-0.9$ &$-2.0$ &$0.9$   &$0.1$  &$0.4$ &$-0.8$ \\
&&$(0.5)$ &$(0.5)$ & $(0.5)$ &$(0.5)$ & $(0.3)$  &$(0.3)$ & $(0.3)$ & $(0.4)$\\
                         & \multirow{2}{*}{MSE}        
& $0.23$ &$0.23$ &$0.27$ &$0.26$ &$0.12$  &$0.11$ &$0.12$ &$0.20$ \\
&&$(0.03)$  &$(0.04)$ & $(0.03)$ & $(0.04)$& $(0.02)$  &$(0.02)$ & $(0.17)$ & $(0.04)$ \\
\hline
\multicolumn{2}{l}{}                  & \multicolumn{4}{l}{true transformation=$\log({x}/({1-x}))$} \\ \hline
\multicolumn{2}{l}{}            & \multicolumn{4}{l}{$n=50$}    & \multicolumn{4}{l}{$n=100$}  \\  \hline
\multirow{4}{*}{VUS}       & \multirow{2}{*}{bias}       
&$-1.5$   &$-4.3$ &$-0.3$ &$-2.1$  &$1.1$ &$-2.4$ &$0.2$ &$-1.1$\\
&&$(0.5)$  &$(0.6)$ &$(0.5)$ & $(0.5)$ & $(0.3)$  & $(0.4)$ & $(0.4)$ & $(0.5)$ \\
                         & \multirow{2}{*}{MSE}        
&$0.30$  &$0.53$ &$0.22$ &$0.30$ & $0.12$  &$0.21$ &$0.14$ &$0.22$ \\ 
&&$(0.04)$ &$(0.07)$ &$(0.03)$ & $(0.05)$ & $(0.02)$ &$(0.03)$ & $(0.03)$ & $(0.04)$ \\
\hline
\end{tabular}
\end{table}

Notice that, BRL has nothing to do with the true transformation since we only consider the rank information given by the data. So it is expected that the simulation result applying to these two different data sets will be almost the same, except for the difference due to sampling.

From Table 1 and Table 2, it is not surprising to find that when the true transformation belongs to the Box-Cox family, the Box-Cox method performs best in terms of both bias and MSE. However, when the true transformation is no longer from this family, the performance of Box-Cox
is not satisfactory, indicating that this method is not robust under the proposed transformation. BRL method has lower bias and MSE of estimates of the parameters  and VUS in this case compared with other methods except for  the parameter $a$.

\subsection{With verification bias}

\subsubsection{Under trinormal assumption}

In this simulation, set $n_0=n_1=n_2=n$, so we generate $N=3n$ data in total with $n$ data from each category. We consider two different trinormal distributions settings. The three distributions are $\N(\mu_1, \sigma_1^2)$, $\N(0,1)$ and $\N(\mu_2, \sigma_2^2)$, after transformation. The parameters $(a,b,c,d)$ and the $\VUS$  can be uniquely determined by  $(\mu_1, \sigma_1, \mu_2, \sigma_2)$ according to (\ref{eq:abcd}) and (\ref{eq:VUS}).

\begin{table}[]
\centering
\caption{Different data settings}
\label{my-label}
\begin{tabular}{ccc}
$(\mu_1, \sigma_1, \mu_2, \sigma_2)$ & $(a,b,c,d)$ & $\VUS$  \\ \hline
$(-1.8, 1.5, 2, 2)$ & $(0.667, -1.2, 0.5, 1)$ & 0.671 \\
$(-2.3, 1, 2, 1)$ & $(1, -2.3, 1, 2)$ & 0.870\\
 \hline
\end{tabular}
\end{table}

Within each simulation setting, two verification mechanisms based on verification models (\ref{eq:threshold}) for threshold value and (\ref{eq:probit}) for probit regression model, are used to generate the  data with verification bias, which somewhat resemble the data we observe in reality. For the former case, $p_1$ is taken to be 0.8 and $p_2$ is taken to be 0.4 in (\ref{eq:threshold}), which will give an average of $48\%$ missing labels. For the latter case, $\beta$ is fixed to be $1$ and $\alpha$ is adjusted to be $0.106, 0.170$ and $0.189$ to achieve $48\%$ missing labels for each of the cases mentioned above. BRL estimates are obtained by 90000 Gibbs samples, (100000 MCMC iterations after 10000 iterations used for burn-in). In total 100 datasets are simulated for the study. For threshold missing scheme (\ref{eq:threshold}), $n$ is set to be 100 and 200, while for probit missing scheme (\ref{eq:probit}), none of the methods considered give results for $n=100$, suggesting that at such a missing rate the number of observations may be insufficient. So we only consider $n=200$ for this missing mechanism.

There are very limited papers dealing with the problem when missing labels are involved in estimating the ROC surface. We compare our method to FI, MSI, IPW and SPE method proposed by To Duc et al. (2016). We use a multivariate logistic model to estimate the true disease rate and a logistic model to estimate verification rate in order to apply those methods. Because these methods can only give estimate at a point, we consider fitting a surface to those points that has the form under trinormality assumption, given by (\ref{eq:roc}). We start with a grid and find the parameters that minimize the sum of squared distance for every point on the grid. The comparison of the estimates are given in the tables below. We also compare those methods in terms of the estimated accuracy of $\VUS$. Notice that for the $\VUS$ of FI, MSI and IPW, we use the calculation of $\VUS$ given in To Duc et al. (2016) instead of just plugging in the parameter estimates in (\ref{eq:VUS}).

\begin{table}[]
\centering
\caption{Bias $(\times 10^{2})$ and MSE $(\times 10^{2})$ for estimates of parameter values under trinormality, with verification bias generated using the threshold model}
\label{my-label}
\begin{tabular}{cccccccccc}
\multicolumn{2}{l}{\multirow{3}{*}{}}  & \multicolumn{4}{l}{$n=100$}     & \multicolumn{4}{l}{$n=200$}   \\ \cmidrule(lr){3-6}\cmidrule(lr){7-10}
\multicolumn{2}{l}{}                  & BRL    & FI   & MSI   & IPW   & BRL    & FI   & MSI  & IPW  \\ \hline
\multicolumn{2}{l}{}                  & \multicolumn{8}{l}{$(a,b,c,d)=(0.667, -1.2, 0.5, 1)$} \\ \hline
\multirow{4}{*}{$a$}       & \multirow{2}{*}{bias}       
&  $-19.5$&$31.4$ &$10.9$  &$-1.9$ &  $-8.2$& $30.8$&$10.0$ &$-3.4$\\
&&$(1.6)$&$(0.4)$&$(1.0)$&$(1.9)$&$(0.8)$&$(0.3)$&$(0.7)$&$(1.0)$\\
                         & \multirow{2}{*}{MSE}        
& $6.5$  &$10.0$ &$2.2$  &$3.5$  &$1.4$   &$9.6$ &$1.5$ &$1.0$ \\
&&$(0.8)$&$(0.2)$&$(0.3)$&$(1.5)$&$(0.2)$&$(0.2)$&$(0.1)$&$(0.2)$\\
\multirow{4}{*}{$b$}       & \multirow{2}{*}{bias}       
& $-4.0$ &$3.0$ &$12.8$  &$-9.3$ &$-0.7$  &$0.7$ &$12.5$ &$-7.6$\\
&&$(2.1)$&$(2.0)$&$(1.9)$&$(1.9)$&$(1.5)$&$(1.5)$&$(1.4)$&$(1.6)$\\
                         & \multirow{2}{*}{MSE}        
& $4.7$  &$4.0$ &$5.0$  &$4.5$  &$2.2$   &$2.3$ &$3.6$ &$3.0$ \\
&&$(1.5)$&$(0.5)$&$(0.6)$&$(1.0)$&$(0.3)$&$(0.3)$&$(0.4)$&$(0.5)$\\
\multirow{4}{*}{$c$}       & \multirow{2}{*}{bias}       
&  $-13.8$&$22.1$&$28.3$  &$12.4$ & $-5.6$ &$25.0$ &$27.1$ &$8.1$ \\
&&$(1.3)$&$(1.6)$&$(2.4)$&$(3.9)$&$(0.8)$&$(1.3)$&$(2.0)$&$(2.2)$\\
                         & \multirow{2}{*}{MSE}        
& $3.5$  &$7.4$ &$13.7$  &$16.8$  & $0.9$  &$7.8$ &$11.5$ &$5.3$ \\
&&$(0.4)$&$(0.6)$&$(2.0)$&$(9.9)$&$(0.1)$&$(0.5)$&$(1.5)$&$(2.0)$\\
\multirow{4}{*}{$d$}       & \multirow{2}{*}{bias}       
&  $-5.3$&$12.1$ &$22.1$  &$16.3$  &$-2.8$  &$18.7$&$24.5$ &$17.8$ \\
&&$(1.9)$&$(2.7)$&$(3.8)$&$(5.6)$&$(1.5)$&$(2.2)$&$(2.7)$&$(3.4)$\\
                         & \multirow{2}{*}{MSE}        
&  $3.8$ &$8.9$ & $18.9$ & $33.8$ & $2.2$ &$8.4$ & $13.2$& $14.7$\\
&&$(0.5)$&$(2.0)$&$(3.9)$&$(17.6)$&$(0.4)$&$(1.0)$&$(2.2)$&$(6.5)$\\
\hline
\multicolumn{2}{l}{}                  & \multicolumn{8}{l}{$(a,b,c,d)=(1, -2.3, 1, 2)$} \\ \hline
\multirow{4}{*}{$a$}       & \multirow{2}{*}{bias}       
& $-10.5$ &$3.1$ &$7.8$ &$7.2$ &$-2.5$ &$0.4$ &$3.7$ &$4.5$   \\
&&$(2.7)$&$(0.8)$&$(1.4)$&$(3.5)$&$(1.8)$&$(0.4)$&$(1.1)$&$(2.9)$\\
                         & \multirow{2}{*}{MSE}
& $8.4$ &$0.7$ &$2.6$ &$12.8$ &$3.3$ &$0.2$ &$1.4$ &$8.6$     \\
&&$(1.2)$&$(0.1)$&$(0.4)$&$(2.4)$&$(0.5)$&$(0.1)$&$(0.3)$&$(2.6)$\\
\multirow{4}{*}{$b$}       & \multirow{2}{*}{bias}       
& $3.0$ &$-7.1$ &$-9.0$ &$-8.8$ &$-1.2$ &$0.6$ &$2.4$ &$-0.1$\\
&&$(3.2)$&$(3.6)$&$(3.8)$&$(5.2)$&$(2.7)$&$(2.4)$&$(2.6)$&$(3.5)$\\
                         & \multirow{2}{*}{MSE}
& $10.3$ &$13.5$ &$15.2$ &$27.9$ &$7.4$ &$5.9$ &$6.6$ &$11.9$     \\
&&$(1.5)$&$(2.2)$&$(2.4)$&$(7.6)$&$(1.1)$&$(1.0)$&$(1.0)$&$(2.9)$\\
\multirow{4}{*}{$c$}       & \multirow{2}{*}{bias}       
& $-7.6$ &$9.9$ &$11.1$ &$18.9$ &$-1.7$ &$11.1$ &$9.4$ &$23.1$   \\
&&$(2.6)$&$(1.8)$&$(3.1)$&$(5.9)$&$(1.6)$&$(1.6)$&$(2.4)$&$(4.6)$\\           
           & \multirow{2}{*}{MSE}
& $7.5$ &$4.1$ &$10.9$ &$37.6$ &$2.5$ &$3.8$ &$6.7$ &$26.7$        \\
&&$(1.3)$&$(0.8)$&$(2.1)$&$(10.0)$&$(0.4)$&$(0.8)$&$(1.2)$&$(5.7)$\\
\multirow{4}{*}{$d$}       & \multirow{2}{*}{bias}       
& $-2.8$ &$18.8$ &$9.9$ &$32.6$ &$0.7$ &$22.0$ &$13.9$ &$48.3$     \\
&&$(4.2)$&$(4.1)$&$(5.0)$&$(8.6)$&$(2.9)$&$(3.5)$&$(4.5)$&$(8.0)$\\           
           & \multirow{2}{*}{MSE}
& $18.0$ &$19.9$ &$25.7$ &$83.1$ &$8.2$ &$16.7$ &$22.1$ &$86.5$     \\    
&&$(4.1)$&$(4.7)$&$(5.4)$&$(22.4)$&$(1.2)$&$(3.8)$&$(3.8)$&$(21.1)$\\
\hline
\end{tabular}
\end{table}

\begin{table}[]
\centering
\caption{Simulation result for the VUS under trinormality, with verification bias generated using the threshold model}
\label{my-label}
\begin{tabular}{rrrrrrr}
\multicolumn{2}{l}{\multirow{3}{*}{}}  & \multicolumn{5}{l}{$n=100$}     \\ \cmidrule(lr){3-7}
\multicolumn{2}{l}{}                  & BRL    & FI   & MSI   & IPW &SPE    \\ \hline
\multicolumn{2}{l}{}                  & \multicolumn{5}{l}{$(a,b,c,d)=(0.667, -1.2, 0.5, 1), \VUS=0.671$} \\ \hline
\multirow{4}{*}{VUS}     & \multirow{2}{*}{bias}       
& $1.9$ &$-1.4$ &$0.1$  & $6.2$ &  $1.4$ \\
&&$(0.4)$&$(0.6)$&$(0.5)$&$(0.5)$&$(0.5)$\\
                         & \multirow{2}{*}{MSE}        
& $0.19$ &$0.34$ &$0.26$ &$0.64$ &$0.26$ \\
&&$(0.03)$&$(0.05)$&$(0.03)$&$(0.06)$&$(0.03)$\\

\hline
\multicolumn{2}{l}{}                  & \multicolumn{5}{l}{$(a,b,c,d)=(1, -2.3, 1, 2), \VUS=0.870$}  \\ \hline
\multirow{4}{*}{VUS}     & \multirow{2}{*}{bias}        
&$0.1$ & $0.8$ & $0.7$ & $2.7$ & $0.5$ \\
&&$(0.3)$&$(0.2)$&$(0.26)$&$(0.3)$&$(0.3)$\\
& \multirow{2}{*}{MSE}         
&$0.08$ &$0.07$ &$0.07$ &$0.14$ &$0.08$     \\ 
&&$(0.01)$&$(0.01)$&$(0.01)$&$(0.01)$&$(0.01)$\\
\hline
\multicolumn{2}{l}{\multirow{3}{*}{}}     & \multicolumn{5}{l}{$n=200$}   \\ \cmidrule(lr){3-7}
\multicolumn{2}{l}{}                  & BRL    & FI   & MSI  & IPW &SPE \\ \hline
\multicolumn{2}{l}{}                  & \multicolumn{5}{l}{$(a,b,c,d)=(0.667, -1.2, 0.5, 1), \VUS=0.671$} \\ \hline
\multirow{4}{*}{VUS}     & \multirow{2}{*}{bias}       
& $0.6$ & $-1.6$  &$-0.2$ &  $6.4$ &  $1.1$\\
&&$(0.3)$&$(0.4)$&$(0.3)$&$(0.3)$& $(0.3)$\\
                         & \multirow{2}{*}{MSE}        
&$0.12$ &$0.17$ &$0.12$ &$0.52$ &$0.12$ \\
&&$(0.02)$&$(0.34)$&$(0.02)$&$(0.05)$&$(0.02)$\\
\hline

\multicolumn{2}{l}{}                  & \multicolumn{5}{l}{$(a,b,c,d)=(1, -2.3, 1, 2), \VUS=0.870$}  \\ \hline
\multirow{4}{*}{VUS}     & \multirow{2}{*}{bias}        
& $0.1$ & $0.4$ & $0.3$ & $2.5$ & $0.3$\\
&&$(0.2)$&$(0.2)$&$(0.2)$&$(0.2)$&$(0.002)$\\
& \multirow{2}{*}{MSE}         
&$0.05$ &$0.05$ &$0.05$ &$0.11$ &$0.06$      \\ 
&&$(<0.01)$&$(0.01)$&$(0.01)$&$(0.01)$&$(0.01)$\\
\hline
\end{tabular}
\end{table}

\begin{table}[]
\centering
\caption{Bias $(\times 10^{2})$ and MSE $(\times 10^{2})$ for estimates of parameter values under trinormality, with verification bias generated using the probit model, $n=200$}
\label{my-label}
\begin{tabular}{rrrrrr}
\multicolumn{2}{l}{}                  & BRL    & FI   & MSI   & IPW     \\ \hline
\multicolumn{2}{l}{}                  & \multicolumn{4}{l}{$(a,b,c,d)=(0.667, -1.2, 0.5, 1)$} \\ \hline
\multirow{4}{*}{$a$}       & \multirow{2}{*}{bias}       
& $-17.3$ &$29.6$ &$22.8$ &$1.8$   \\
&&$(1.0)$&$(0.2)$&$(0.5)$&$(1.9)$\\
                         & \multirow{2}{*}{MSE}         
                         & $4.1$ &$8.8$ &$5.5$ &$3.6$ \\
                         &&$(0.4)$&$(0.1)$&$(0.2)$&$(0.9)$\\
\multirow{4}{*}{$b$}       & \multirow{2}{*}{bias}        
& $-11.7$ &$40.7$ &$42.2$ &$33.6$  \\
&&$(1.3)$&$(2.7)$&$(2.6)$&$(2.6)$\\
                         & \multirow{2}{*}{MSE}         
                         & $2.9$ &$23.7$ &$24.3$ &$18.0$ \\
                         &&$(0.4)$&$(2.7)$&$(2.7)$&$(2.7)$\\
\multirow{4}{*}{$c$}       & \multirow{2}{*}{bias}        
& $-9.6$ &$35.6$ &$53.9$ &$8.1$ \\
&&$(1.0)$&$(2.8)$&$(3.7)$&$(1.6)$\\
                         & \multirow{2}{*}{MSE}         
                         & $1.9$ &$20.3$ &$42.4$ &$3.2$  \\
                         &&$(0.2)$&$(5.9)$&$(4.5)$&$(0.6)$\\
\multirow{4}{*}{$d$}       & \multirow{2}{*}{bias}        
& $-12.5$ &$61.2$ &$69.6$ &$-1.5$ \\
&&$(1.8)$&$(6.1)$&$(6.1)$&$(3.1)$\\
                         & \multirow{2}{*}{MSE}         
                         & $4.9$ &$73.9$ &$8.5$ &$9.3$\\ 
                         &&$(0.6)$&$(27.1)$&$(13.2)$&$(1.6)$\\
                         \hline
\multicolumn{2}{l}{}                  & \multicolumn{4}{l}{$(a,b,c,d)=(1, -2.3, 1, 2)$} \\ \hline
\multirow{4}{*}{a}       & \multirow{2}{*}{bias}        
&$-10.3$ &$2.0$ &$4.1$ &$-1.5$    \\
&&$(2.7)$&$(0.6)$&$(1.9)$&$(3.3)$\\
& \multirow{2}{*}{MSE}         
&$8.3$ &$0.4$ &$3.7$ &$10.6$  \\
&&$(1.0)$&$(0.1)$&$(0.6)$&$(2.6)$\\
\multirow{4}{*}{b}       & \multirow{2}{*}{bias}        
&$-6.6$ &$-11.5$ &$-7.9$ &$32.1$ \\
&&$(3.2)$&$(3.4)$&$(4.2)$&$(3.7)$\\
& \multirow{2}{*}{MSE}         
& $10.8$ &$12.8$ &$18.3$ &$23.8$ \\
&&$(3.3)$&$(2.2)$&$(3.0)$&$(2.5)$\\
\multirow{4}{*}{c}       & \multirow{2}{*}{bias}        
& $-9.5$ &$8.9$ &$1.7$ &$-3.1$\\
&&$(1.5)$&$(2.1)$&$(2.7)$&$(2.1)$\\
& \multirow{2}{*}{MSE}         
& $3.1$ &$5.1$ &$7.4$ &$4.6$ \\
&&$(0.5)$&$(1.6)$&$(1.8)$&$(0.7)$\\
\multirow{4}{*}{d}       & \multirow{2}{*}{bias}       
& $-12.7$ &$18.3$ &$0.6$ &$-13.4$ \\
&&$(2.3)$&$(4.2)$&$(5.0)$&$(4.4)$\\
& \multirow{2}{*}{MSE}
& $6.9$ &$20.8$ &$25.2$ &$20.6$ \\                                  
&&$(1.0)$&$(5.8)$&$(6.4)$&$(3.7)$\\ \hline      

\end{tabular}
\end{table}

\begin{table}[]
\centering
\caption{Bias $(\times 10^{2})$ and MSE $(\times 10^{2})$ for estimates of VUS under trinormality, with verification bias generated using the probit model, $n=200$}
\label{my-label}
\begin{tabular}{ccccccc}
\multicolumn{2}{l}{}                  & BRL    & FI   & MSI   & IPW &SPE    \\ \hline
\multicolumn{2}{l}{}                  & \multicolumn{5}{l}{$(a,b,c,d)=(0.667, -1.2, 0.5, 1)$, VUS=0.671} \\ \hline
\multirow{4}{*}{VUS}     & \multirow{2}{*}{bias}       
& $1.2$ & $-5.4$ & $-5.3$ & $-8.2$ & $-6.4$\\
&&$(0.4)$&$(0.7)$&$(0.7)$&$(1.2)$&$(1.1)$\\
                         & \multirow{2}{*}{MSE}        
                         & $0.16$ &$0.79$ &$0.78$ &$2.1$ &$1.5$ \\ 
                         &&$(0.02)$&$(0.14)$&$(0.15)$&$(0.42)$&$(0.29)$\\
                         \hline

\multicolumn{2}{l}{}                  & \multicolumn{5}{l}{$(a,b,c,d)=(1, -2.3, 2, 1)$, VUS=0.870} \\ \hline
\multirow{4}{*}{VUS}     & \multirow{2}{*}{bias}       
& $0.2$ & $0.5$ & $0.5$ & $-2.8$ & $0.6$\\
&&$(0.2)$&$(0.2)$&$(0.2)$&$(0.6)$&$(0.3)$\\
                         & \multirow{2}{*}{MSE}        
                         & $0.06$ &$0.05$ &$0.05$ & $0.49$ &$0.11$\\
                         && $(0.01)$&$(0.01)$&$(0.01)$&$(0.11)$&$(0.02)$\\
                         \hline                         
\end{tabular}
\end{table}

From Tables 4 and 6, we can see that BRL performs consistently better in term of accuracy in almost all of the cases considered above for parameter estimates. The estimates of the parameter $a$ from BRL sometimes have more bias and larger MSE compared with other methods. The estimates of the parameter $b$ sometimes have more bias but is more accurate in terms of MSE. The estimates of the parameters $c$ and $d$ from BRL are more accurate in terms of both bias and MSE. From Tables 5 and 7, we see that BRL may have larger bias but always has the smallest MSE on estimating VUS.
To sum up, BRL has the best performance overall. It is also worth mentioning that when the sample size increases, accuracy of all methods considered improved, as expected. The improvement is the biggest for BRL.

\subsubsection{Departure from trinormality assumption}

It is important to study the performance of our method when the data does not satisfy the trinormality assumption. Here we generate $(X_1, ..., X_{n_0})$ independently from $\mathrm{Beta}(3,5)$, $(Y_1, ..., Y_{n_1})$ independently from $\mathrm{Beta}(2,2)$, $(Z_1, ..., Z_{n_2})$ independently from $\mathrm{Beta}(5,3)$, where $n_0=n_1=n_2=n=200$. The corresponding $\VUS=0.35815$ and 100 simulated data sets are used in the study. For the verification bias models, we use $p_1=0.8$ and $p_2=0.4$ in (\ref{eq:threshold}), and $\alpha=0.01$ and $\beta=0.07$ in (\ref{eq:probit}), both of them give missing rates  approximately $48\%$. BRL estimates are obtained by 90000 Gibbs samples, (100000 MCMC iterations after 10000 iterations used for burn-in).

\begin{table}[]
\centering
\caption{Bias $(\times 10^{2})$ and MSE $(\times 10^{2})$ for estimates of VUS when departing from the trinormality assumption, $n=200$}
\label{my-label}
\begin{tabular}{ccccccc}
\multicolumn{2}{l}{\multirow{2}{*}{}} & \multicolumn{5}{l}{Threshold} \\ \cmidrule(lr){3-7} 
\multicolumn{2}{l}{}                  & BRL    & FI   & MSI   & IPW &SPE  \\ \hline
\multirow{4}{*}{VUS}     & \multirow{2}{*}{bias}       
& $-0.3$ & $4.6$ &  $2.4$ &  $1.2$ & $-1.4$ \\
&&$(0.2)$ &$(0.2)$&$(0.3)$ &$(0.3)$ &$(0.3)$ \\
                         & \multirow{2}{*}{MSE}         
& $0.03$ &$0.28$ &$0.12$ &$0.13$ &$0.13$ \\ 
&&$(0.01)$&$(0.02)$&$(0.02)$&$(0.02)$&$(0.02)$\\\hline
\multicolumn{2}{l}{\multirow{2}{*}{}} & \multicolumn{5}{l}{Probit} \\ \cmidrule(lr){3-7} 
\multicolumn{2}{l}{}                 & BRL    & FI   & MSI   & IPW &SPE  \\ \hline
\multirow{4}{*}{VUS}     & \multirow{2}{*}{bias}       
& $-1.6$ & $4.8$ & $2.5$ & $0.1$ & $0.1$\\
&&$(0.2)$ &$(0.3)$&$(0.3)$&$(0.3)$&$(0.3)$\\
                         & \multirow{2}{*}{MSE}         
 & $0.08$ &$0.30$ &$0.14$ &$0.10$ &$0.10$\\ 
&& $(0.01)$&$(0.03)$&$(0.02)$&$(0.01)$&$(0.01)$\\\hline
\end{tabular}
\end{table}

We see from Table 8 that BRL estimates are more accurate than other methods for the VUS especially in term of the  MSE and under the threshold missing mechanism. Under the probit model, BRL has slightly more bias and larger MSE compared with IPW and SPE but the difference is insignificant. The simulation results show that BRL is unexpectedly robust against departure from trinormality. 

\subsubsection{Departure from the MAR assumption}

When the MAR assumption failed, the verification probability is no longer independent of disease status conditional on observed values. Thus the model we have in (\ref{eq:vb}) is no longer correct. More precisely,  we will have
\begin{equation}
\begin{aligned}
\P(L_i \neq 3|Q_i, D_i=0)=g_0(Q_i), \\
\P(L_i \neq 3|Q_i, D_i=1)=g_1(Q_i), \\
\P(L_i \neq 3|Q_i, D_i=2)=g_2(Q_i),
\end{aligned}
\end{equation}
then as in the proof of Lemma 2, it follows that
\begin{equation}
\begin{aligned}
\P(D_i=0|& Q_i=t, L_i=3) = \P(D_i=0| Q_i=t, L_i \neq 3)\\
&=\dfrac{\lambda_0 g_0(t) \phi_{(\mu_1, \sigma_1)}(t )}{\lambda_0 g_0(t) \phi_{(\mu_1, \sigma_1)}(t )+\lambda_1 g_1(t) \phi(t )+\lambda_2 g_2(t) \phi_{(\mu_2, \sigma_2)}(t )}, \\
\P(D_i=1|& Q_i=t, L_i=3) = \P(D_i=1| Q_i=t, L_i \neq 3)\\
&=\dfrac{\lambda_1 g_1(t) \phi(t )}{\lambda_0 g_0(t) \phi_{(\mu_1, \sigma_1)}(t )+\lambda_1 g_1(t) \phi(t )+\lambda_2 g_2(t) \phi_{(\mu_2, \sigma_2)}(t )},,\\
\P(D_i=2|& Q_i=t, L_i=3) = \P(D_i=2| Q_i=t, L_i \neq 3)\\
&=\dfrac{\lambda_2 g_2(t) \phi_{(\mu_2, \sigma_2)}(t )}{\lambda_0 g_0(t) \phi_{(\mu_1, \sigma_1)}(t )+\lambda_1 g_1(t) \phi(t )+\lambda_2 g_2(t) \phi_{(\mu_2, \sigma_2)}(t )}.
\end{aligned}
\end{equation}

The true normal distributions considered here are $\N(\mu_1, \sigma_1^2)$, $\N(0,1)$ and $\N(\mu_2, \sigma_2^2)$, where $(\mu_1, \sigma_1, \mu_2, \sigma_2)=(-1.8, 1.5, 2, 2)$. We consider two cases when the MAR assumption fails. Let $\lambda_0=\lambda_1=\lambda_2=1/3$, to achieve 48\% verification probability, i.e. $\P(L_i \neq 3)=0.48$, let $\P(L_i \neq 3| D_i=0)=0.72$, $\P(L_i \neq 3| D_i=1)=0.48$ and $\P(L_i \neq 3| D_i=2)=0.24$. Consider two cases here.

\begin{enumerate}
\item  Let $g(Q; p_1, p_2)=
\begin{cases}
1, \ \ &\mbox{if } Q>Q_{(p_{1}N)}, \\ 
p_{2}, \ \ &\mbox{if }Q\leq Q_{(p_{1}N)}.  
\end{cases}$

The verification model for healthy group is $g_0(Q)=g(Q; 0.8, 0.1)$; 
the verification model for level-1 disease is $g_1(Q)=g(Q; 0.6, 0.2)$;
the verification model for level-2 disease is $g_2(Q)=g(Q; 0.4, 0.4)$. 

\item Let $g(Q; \alpha, \beta)=\Phi(\alpha+\beta Q)$.

The verification model for healthy group is $g_0(Q)=g(Q; 0.217, 0.5)$; 
the verification model for level-1 disease is $g_1(Q)=g(Q; 0.052, 0.3)$;
the verification model for level-2 disease is $g_2(Q)=g(Q; 0.334, 0.2)$. 

\end{enumerate}

Here $n_0=n_1=n_2=n$ is set to be 200. Again we simulate 100 data sets for the study, and BRL estimates are obtained by 90000 Gibbs samples, (100000 MCMC iterations after 10000 iterations used for burn-in). 

\begin{table}[]
\centering
\caption{Bias $(\times 10^{2})$ and MSE $(\times 10^{2})$ for estimates of parameter estimates when departing from the MAR assumption, $n=200$}
\label{my-label}
\begin{tabular}{rrrrrrrrrr}
\multicolumn{2}{l}{\multirow{2}{*}{}} & \multicolumn{4}{c}{Threshold} & \multicolumn{4}{c}{Probit} \\ \cmidrule(lr){3-6} \cmidrule(lr){7-10}
\multicolumn{2}{l}{}                  & BRL    & FI   & MSI   & IPW  & BRL    & FI   & MSI   & IPW   \\ \hline
\multirow{2}{*}{$a$}       & \multirow{2}{*}{bias} 
& $-12.3$ &$30.3$ &$21.6$ &$-8.9$ &$-10.6$ &$31.6$ &$20.1$ &$-1.7$ \\
&&$(1.3)$&$(0.3)$&$(0.6)$&$(1.5)$&$(1.0)$&$(0.3)$&$(0.7)$&$(1.2)$\\
& \multirow{2}{*}{MSE}
& $3.1$ &$9.2$ &$5.1$ &$3.2$ &$2.9$ &$10.0$ &$4.5$ &$1.3$\\
&&$(0.4)$&$(0.2)$&$(0.3)$&$(0.4)$&$(0.3)$&$(0.2)$&$(0.3)$&$(0.2)$\\
\multirow{2}{*}{$b$}       & \multirow{2}{*}{bias}      
& $1.8$ &$38.0$ &$41.7$ &$-19.3$ &$5.5$ &$30.4$ &$35.3$ &$37.2$\\
&&$(1.4)$&$(1.8)$&$(1.7)$&$(3.5)$&$(1.5)$&$(1.9)$&$(1.7)$&$(1.7)$\\
& \multirow{2}{*}{MSE}    
& $2.1$ &$17.7$ &$20.3$ &$15.9$ &$2.6$ &$12.9$ &$15.4$ &$16.9$ \\
&&$(0.3)$&$(1.4)$&$(1.4)$&$(2.1)$&$(0.4)$&$(1.3)$&$(1.3)$&$(1.5)$\\
\multirow{2}{*}{$c$}       & \multirow{2}{*}{bias}      
& $-23.4$ &$30.4$ &$37.1$ &$-15.0$ &$-11.8$ &$28.8$ &$34.3$ &$5.9$\\
&&$(0.7)$&$(1.7)$&$(3.2)$&$(0.9)$&$(0.8)$&$(1.3)$&$(2.4)$&$(1.4)$\\
& \multirow{2}{*}{MSE}
& $6.0$ &$12.1$ &$24.2$ &$3.0$ & $2.1$ &$10.1$ &$17.4$ &$2.4$ \\
&&$(3.7)$&$(1.4)$&$(5.6)$&$(0.3)$&$(0.2)$&$(0.7)$&$(1.9)$&$(0.6)$\\
\multirow{2}{*}{$d$}       & \multirow{2}{*}{bias}       
& $-43.3$ &$37.2$ &$51.5$ &$-50.3$ & $-23.0$ &$19.6$ &$24.7$ &$-19.5$\\
&&$(1.5)$&$(3.0)$&$(6.6)$&$(3.8)$&$(1.5)$&$(2.3)$&$(3.1)$&$(1.9)$\\
& \multirow{2}{*}{MSE}
& $20.9$ &$22.8$ &$69.3$ &$39.8$ &$7.6$ &$9.3$ &$15.6$ &$7.5$\\   
&&$(1.3)$&$(3.2)$&$(23.7)$&$(5.1)$&$(0.7)$&$(1.1)$&$(2.6)$&$(0.8)$\\ \hline
\end{tabular}
\end{table}

\begin{table}[]
\centering
\caption{Bias $(\times 10^{2})$ and MSE $(\times 10^{2})$ for estimates of the VUS when departing from the MAR assumption, $n=200$}
\label{my-label}
\begin{tabular}{rrrrrrr}
\multicolumn{2}{l}{\multirow{2}{*}{}} & \multicolumn{5}{c}{Threshold} \\ \cmidrule(lr){3-7}
\multicolumn{2}{l}{}                  & BRL    & FI   & MSI   & IPW &SPE  \\ \hline
\multirow{4}{*}{VUS}     & \multirow{2}{*}{bias}        
& $-8.3$ &$-6.8$ &$-6.1$ & $-6.5$ & $-5.4$\\
&&$(0.3)$&$(0.5)$&$(0.5)$&$(1.2)$&$(1.7)$\\
                         & \multirow{2}{*}{MSE}         
                         & $0.79$ &$0.70$ &$0.59$ &$1.89$ &$3.01$ \\ 
                         &&$(0.06)$&$(0.07)$&$(0.06)$&$(0.37)$&$(0.78)$\\
                         \hline
\multicolumn{2}{l}{\multirow{2}{*}{}} & \multicolumn{5}{c}{Probit} \\ \cmidrule(lr){3-7}
\multicolumn{2}{l}{}                  & BRL    & FI   & MSI   & IPW &SPE  \\ \hline
\multirow{4}{*}{VUS}     & \multirow{2}{*}{bias}        
& $-4.5$ & $-7.9$ & $-6.9$ & $-9.8$ & $-7.7$\\
&&$(0.4)$  &$(0.4)$ &$(0.4)$ &$(0.4)$ &$(0.4)$\\
                         & \multirow{2}{*}{MSE}         
                         &$0.34$ &$0.80$ &$0.63$ &$1.13$ &$0.77$\\ 
                         &&$(0.04)$  &$(0.08)$&$(0.06)$&$(0.09)$&$(0.08)$\\
                         \hline
\end{tabular}
\end{table}

It is expected that the bias and MSE are larger compared with the situation under MAR mechanism, which is indeed shown in Tables 9 and 10. We also see that still BRL has more accurate estimates of parameters. For VUS, under the threshold model, BRL has slightly more bias and larger MSE compared with FI and MSI, but the difference is insignificant, while under probit model BRL still has the best performance. Thus BRL is robust against departure from the MAR assumption.

\section{Real data analysis}

As an illustration of the application of our method, we apply this to a real data set on the diagnosis of epithelial ovarian cancer (EOC). This data  from the SPORE/Early Detection
Network/Prostate, Lung, Colon, and Ovarian Cancer Ovarian Validation
Study is available to public.\footnote{https://edrn.nci.nih.gov/protocols/119-spore-edrn-pre-plco-ovarian-phase-ii-validation} This data is also partially available in R-package bcROCsurface by To Duc (2017) as an implementation of their  bias-corrected methods for the ROC surface. We follow their setting by using this data set and compare their methods.

This data set is used for evaluating the performance of several biomarkers in diagnosis of EOC. There are three classes included in this data set, i.e., benign disease, early stage and late stage cancer. We want to know which biomarkers are the most informative on their own in prediction on these three classes. The two particular biomarkers we compare in this example are CA125 and CA153, measured at Havard laboratories. We have 134 patients with benign disease, 67 in early stage  and 77 in late stage. Since all of our samples are verified, there is no verification bias issue here. So we will apply our BRL method on without verification bias setting and then compare this to empirical estimates.

To illustrate the verification bias case, To Duc et al. (2016) simulated a verification process, i.e., $\P(L_i = 3) = 0.05 + 0.35 \mathbbm{1}(\mbox{CA}125_i > 0.87) + 0.25 \mathbbm{1}(\mbox{CA}153_i > 0.3) + 0.35 \mathbbm{1}(\mbox{Age}_i > 45)$. This process leads to 63.4\% patients selected to undergo disease verification. We compare BRL with FI, MSI, IPW and SPE, where a logistic model is used to estimate the true disease rate and a logistic model to estimate verification rate as well.

First we look at the ROC surface of CA125. In Table 11 we give the estimates of parameters of the ROC surface for BRL applying on the full data and then to the data obtained under verification bias. It can be seen from this table that the parameter estimates using BRL in those two cases are similar, with $b$ having the biggest difference of around $0.2$. Also the standard deviation of parameter estimates increased from the full data to the data under verification bias, which is expected. 

\begin{table}[]
\centering
\caption{Parameter estimates from BRL methods for CA125}
\label{my-label}
\begin{tabular}{lrccrcc}
\multicolumn{1}{l}{\multirow{2}{*}{}} & \multicolumn{3}{c}{full} & \multicolumn{3}{c}{with verification bias} \\ \cmidrule(lr){2-4} \cmidrule(lr){5-7}
\multicolumn{1}{l}{}                  & estimate &sd &95\% C.I. &estimates &sd  &95\% C.I.  \\ \hline
$a$ & 1.151 &0.168 &$[0.822, 1.481]$ & 1.176 & 0.228 &$[0.730, 1.622]$\\
$b$ & $-1.406$ &0.193 &$[-1.784, -1.028]$&$-1.217$ &$0.236$ &$[-1.680, 0.236]$\\
$c$ & 0.821 &0.124 &$[0.578, 1.063]$ &0.930 &0.168 &$[0.600, 1.260]$\\
$d$ & 0.723 &0.166  &$[0.398, 1.049]$ &0.774 &0.230 &$[0.323, 1.225]$\\ \hline
\end{tabular}
\end{table}

In Table 12, we give the estimates of the VUS by empirical and BRL methods using the full data and BRL, FI, MSI, IPW and SPE using the data with verification bias (BRL method applied on the  full data is denoted by BRL full while BRL method applied on the data with verification bias is denoted by BRL vb). The BRL methods are based on 250000 iterations as the chain has 300000 iterations after 50000 iterations as burn-in. The estimates of the VUS are similar using all of those methods, with BRL applying to data with verification bias has the smallest value 0.511, and the empirical method applied to the full dataset has the largest value 0.566. Also it seems that the BRL method applied to the data with verification bias has the largest standard deviation.

\begin{table}[]
\centering
\caption{VUS estimates for CA125}
\label{my-label}
\begin{tabular}{lccc}
\hline
 & estimated VUS  &sd &95\% C.I. \\ \hline
Full & 0.566 &0.037 & $[0.495, 0.638]$ \\
BRL full & 0.545 &0.040 & $[0.467, 0.623]$ \\
BRL vb & 0.511 & 0.048 &$[0.417, 0.605]$ \\
FI & 0.515 &0.040 &$[0.436, 0.594]$ \\
MSI &0.518 &0.042 &$[0.437, 0.600]$ \\
IPW &0.550 &0.042 &$[0.469, 0.631]$ \\
SPE &0.558 &0.044 &$[0.471, 0.645]$ \\ \hline
\end{tabular}
\end{table}

To better compare the fitted ROC surface using all those methods, we also plot out the  ROC surface.

\begin{figure}
\centering
\includegraphics[width=1.1\textwidth]{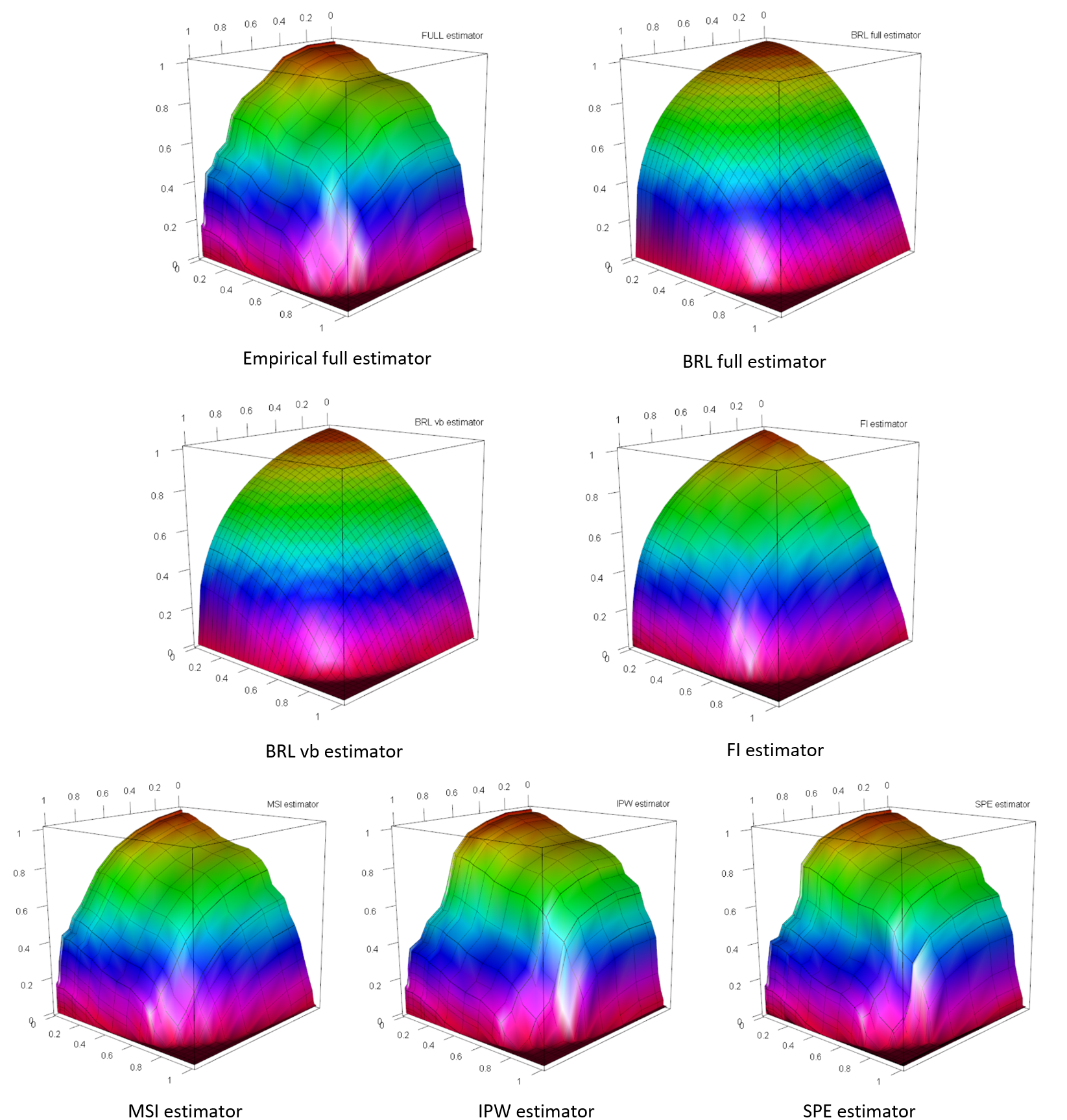}
\caption{\label{fig:CA125}Estimated ROC surfaces for CA125}
\end{figure}

We can from Figure 1 that since the BRL method assumes trinormality while other methods do not, BRL generates a smooth surface. Often a smooth surface is desirable as it is more realistic. The ROC surface fitted using FI looks very similar to the one fitted by the BRL method. The other three estimated surfaces using the data with verification bias (i.e., MSI, IPW, SPE), look more like the empirical estimator.

For the ROC surface for CA153, the estimates with full data using BRL methods applying on the full dataset and on the data with  verification bias are mostly similar, having difference within $0.2$. Also the posterior standard deviation increases under verification bias.

\begin{table}[]
\centering
\caption{Parameter estimates from BRL methods for CA153}
\label{my-label}
\begin{tabular}{crccrcc}
\multicolumn{1}{l}{\multirow{2}{*}{}} & \multicolumn{3}{c}{full} & \multicolumn{3}{c}{with verification bias} \\ \cmidrule(lr){2-4} \cmidrule(lr){5-7}
\multicolumn{1}{l}{}                  & estimate &sd &95\% C.I. &estimates &sd  &95\% C.I.  \\ \hline
a & 1.356 &0.153 & $[1.205, 1.508]$ &1.272 &0.229 &$[0.823, 1.721]$\\
b & $-0.407$ &0.169 &$[-0.574, -0.240]$ &$-0.407$ &0.220 &$[-0.838, 0.024]$\\
c & 0.907 &0.140 & $[0.769, 1.046]$ &0.787 &0.169 &$[0.456, 1.118]$\\
d & 0.816 &0.164 &$[0.653, 0.978]$ &0.699 &0.218 &$[0.272, 1.126]$\\ \hline
\end{tabular}
\end{table}

The estimates of VUS using all the methods are shown in Table 14. The estimates are very similar, with FI giving the largest estimate   while IPW giving the smallest. 

\begin{table}[]
\centering
\caption{VUS estimates for CA153}
\label{my-label}
\begin{tabular}{lccc}
\hline
 & estimated VUS &sd &95\% C.I. \\ \hline
Full & 0.356 &0.037 &[0.284, 0.427] \\
BRL full & 0.363 &0.033 & [0.330, 0.396]\\
BRL vb & 0.360 &0.045 & [0.314, 0.405]\\
FI & 0.393 &0.041 &[0.313, 0.474] \\
MSI &0.385 &0.042 &[0.302, 0.468] \\
IPW &0.349 &0.048 &[0.256, 0.443] \\
SPE &0.360 &0.053 &[0.256, 0.464] \\ \hline
\end{tabular}
\end{table}

\begin{figure}
\centering
\includegraphics[width=1.1\textwidth]{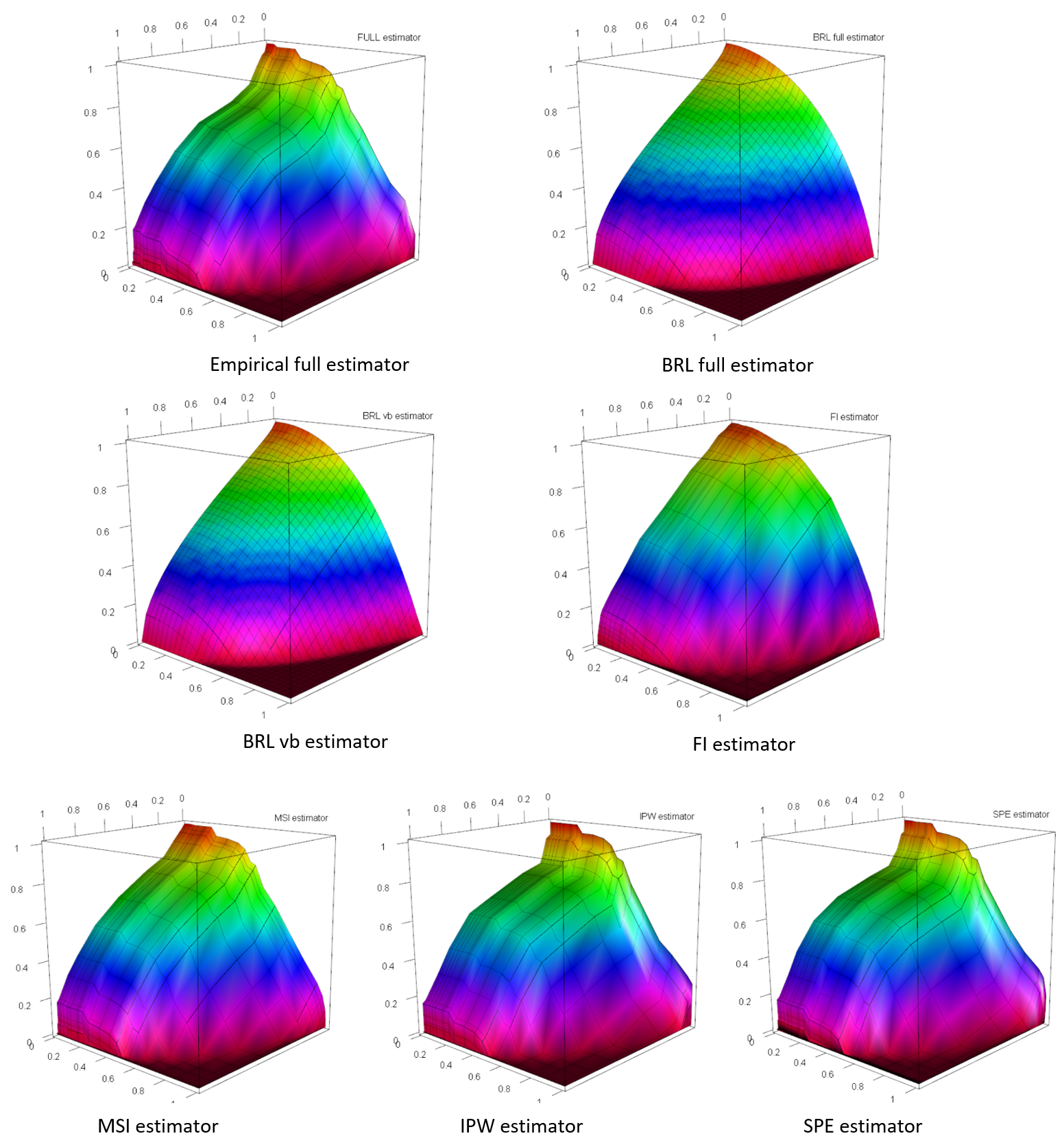}
\caption{\label{fig:CA153}Estimated ROC surfaces for CA153}
\end{figure}

The fitted ROC surface is shown in Figure 2, which gives a qualitatively similar result, i.e., 
the ROC surface fitted using FI estimator is smoother and looks  very much like surface fitted by BRL. The other three estimated surface on data with verification bias (i.e., MSI, IPW, SPE) look like the empirical full estimator.

No matter which estimation method we choose to use, the estimated VUS seems to suggest the same result, namely, the 95\% C.I. for CA125 is always above the 95\% C.I. for CA 153. That is,  CA125 is more accurate in predicting cancer than CA153. A formal comparison would involve a hypothesis testing of two AUC values, which will not be discussed further here.

Youden Index (YI) is another widely used measure of the accuracy of a classifier (Youden, 1950).  We estimate the Younden Index based on different the estimated ROC surface. For BRL full and BRL vb method, the trinormality assumption gives an explicit formula (Luo and Xiong, 2013). For other methods, numerical calculation will be used. The posterior standard deviations of BRL full and BRL vb are calculated in Monte Carlo simulation, while the standard deviations for others will be calculated using the bootstrap method with 1000 resamples.

\begin{table}[]
\centering
\caption{Youden Index estimates for CA125}
\label{my-label}
\begin{tabular}{lccc}
\hline
 & YI estimate &sd &95\% C.I. \\ \hline
Full & 0.901 &0.076 & $[0.751, 1.051]$\\
BRL full & 0.813 & 0.070 & $[0.675, 0.951]$ \\
BRL vb & 0.753 & 0.081 & $[0.594, 0.912]$\\
FI & 0.792 &0.073 &$[0.649, 0.935]$\\
MSI &0.811 & 0.081 & $[0.653, 0.969]$\\
IPW &0.899 & 0.100& $[0.703, 1.095]$\\
SPE &0.924 & 0.102 & $[0.724, 1.124]$\\ \hline
\end{tabular}
\end{table}

\begin{table}[]
\centering
\caption{Youden Index estimates for CA153}
\label{my-label}
\begin{tabular}{lccc}
\hline
 & YI estimate &sd &95\% C.I. \\ \hline
Full & 0.592 & 0.078 & $[0.439, 0.745]$ \\
BRL full & 0.513 & 0.055 & $[0.405, 0.621]$ \\
BRL vb & 0.502 &0.074 & $[0.356, 0.647]$ \\
FI & 0.543 & 0.076 & $[0.394, 0.692]$\\
MSI &0.619 & 0.088 & $[0.447, 0.791]$ \\
IPW &0.638 & 0.099 & $[0.445, 0.831]$\\
SPE &0.676 & 0.101 & $[0.479, 0.873]$\\ \hline
\end{tabular}
\end{table}

Estimates of the Youden Index given by different methods are slightly different from each other, with IPW and SPE mostly giving higher estimates and higher standard deviation, The BRL method applied to the full data or data under verification bias give lower estimation result and smaller standard deviation. However, all of them consistently suggesting that CA 125 is a better classifier for cancer, which coincides with the conclusion we get from VUS estimation.

\section{Appendix}

\begin{proof} \textbf{of Lemma \ref{lemma2}}
According to Bayes' theorem, we have
\begin{equation}
\P(D_i=0|Q_i=t, L_i=3)=\dfrac{\P(D_i=0)f_{Q_i}(t|D_i=0)\P(L_i=3|Q_i=t, D_i=0)}{\sum_{d=0}^2\{\P(D_i=d)f_{Q_i}(t|D_i=d)\P(L_i=3|Q_i=t, D_i=d)\}}.
\end{equation}

Here $\P(L_i=3|Q_i=t, D_i=d)=1-g(t)$ under the MAR model as specified in (\ref{eq:vb}) for $d=0,1,2$, plugging in the above equation, we get 
\begin{equation*}
\begin{aligned}
\P(D_i=0| Q_i=t, L_i=3) &= \dfrac{\lambda_0 \phi_{(\mu_1, \sigma_1)}(t)(1-g(t))}{\lambda_0 \phi_{(\mu_1, \sigma_1)}(t)(1-g(t))+\lambda_1 \phi(t)(1-g(t))+\lambda_2 \phi_{(\mu_2, \sigma_2)}(t)(1-g(t))} \\
&= \dfrac{\lambda_0 \phi_{(\mu_1, \sigma_1)}(t)}{\lambda_0 \phi_{(\mu_1, \sigma_1)}(t)+\lambda_1 \phi(t)+\lambda_2 \phi_{(\mu_2, \sigma_2)}(t)},
\end{aligned}
\end{equation*}
and $$\P(L_i=3)=\int \P(L_i=3|Q_i=t)f_{Q_i}(t)dt=\int (1-g(t))\{\lambda_0 \phi_{(\mu_1, \sigma_1)}(t)+\lambda_1 \phi(t)+\lambda_2 \phi_{(\mu_2, \sigma_2)}(t)\}dt.$$

Similarly, we can prove that
$$
\P(D_i=0| Q_i=t, L_i \neq 3)= \dfrac{\lambda_0 \phi_{(\mu_1, \sigma_1)}(t)}{\lambda_0 \phi_{(\mu_1, \sigma_1)}(t)+\lambda_1 \phi(t)+\lambda_2 \phi_{(\mu_2, \sigma_2)},(t)}
$$
and $$\P(L_i \neq 3)=\int \P(L_i \neq 3|Q_i=t)f_{Q_i}(t)dt=\int g(t)\{\lambda_0 \phi_{(\mu_1, \sigma_1)}(t)+\lambda_1 \phi(t)+\lambda_2 \phi_{(\mu_2, \sigma_2)}(t)\}dt.$$

\end{proof}

\begin{proof}
\textbf{of Lemma \ref{lemma3}}
According to Bayes' theorem, we have
\begin{equation*}
\begin{aligned}
f_Q(t|L \neq 3) = &\dfrac{\P(L=0 \ \mbox{or} \ 1 \ \mbox{or} \ 2 |Q=t)f_Q(t)}{\int \P(L=0 \ \mbox{or} \ 1 \ \mbox{or} \ 2 |Q=s)f_Q(s) ds} \\
= & \dfrac{g(t)\{(1-\lambda_1-\lambda_2)\phi_{(\mu_1, \sigma_1)}(t)+\lambda_1\phi(t)+\lambda_2 \phi_{(\mu_2, \sigma_2)}(t)\}}{\int g(s)\{(1-\lambda_1-\lambda_2)\phi_{(\mu_1, \sigma_1)}(s)+\lambda_1\phi(s)+\lambda_2 \phi_{(\mu_2, \sigma_2)}(s)\} ds} \\
= & \dfrac{g(t) \phi_{(\mu_1, \sigma_1)}(t)}{\int g(s) \phi_{(\mu_1, \sigma_1)}(s) ds} \times \dfrac{(1-\lambda_1-\lambda_2)\int g(s) \phi_{(\mu_1, \sigma_1)}(s) ds}{\int g(s)\{(1-\lambda_1-\lambda_2)\phi_{(\mu_1, \sigma_1)}(s)+\lambda_1\phi(s)+\lambda_2 \phi_{(\mu_2, \sigma_2)}(s)\} ds} \\
& +\dfrac{g(t) \phi(t)}{\int g(s)\phi(s) ds} \times \dfrac{\lambda_1 \int g(s)\phi(s) ds}{\int g(s)\{(1-\lambda_1-\lambda_2)\phi_{(\mu_1, \sigma_1)}(s)+\lambda_1\phi(s)+\lambda_2 \phi_{(\mu_2, \sigma_2)}(s)\} ds} \\
& +\dfrac{g(t) \phi_{(\mu_2, \sigma_2)}(t)}{\int g(s)\phi_{(\mu_2, \sigma_2)}(s) ds} \times \dfrac{\lambda_2 \int g(s)\phi_{(\mu_2, \sigma_2)}(s) ds}{\int g(s)\{(1-\lambda_1-\lambda_2)\phi_{(\mu_1, \sigma_1)}(s)+\lambda_1\phi(s)+\lambda_2 \phi_{(\mu_2, \sigma_2)}(s)\} ds} \\
= & (1-\lambda_{1(\mu_1, \sigma_1, \mu_2, \sigma_2, g)}^*-\lambda_{2(\mu_1, \sigma_1, \mu_2, \sigma_2, g)}^*)\phi^*_{(\mu_1, \sigma_1, g)}(t) \\
&+ \lambda_{1(\mu_1, \sigma_1, \mu_2, \sigma_2, g)}^* \phi^*_{(g)}(t) + \lambda_{2(\mu_1, \sigma_1, \mu_2, \sigma_2, g)}^*\phi^*_{(\mu_2, \sigma_2, g)}(t),
\end{aligned}
\end{equation*}
where $\lambda_{1(\mu_1, \sigma_1, \mu_2, \sigma_2, g)}^*$, $\lambda_{2(\mu_1, \sigma_1, \mu_2, \sigma_2, g)}^*$ and $\phi^*_{(\mu, \sigma, g)}(t)$ are defined in (\ref{def:lambda}).

\end{proof}


\begin{proof}
\textbf{of Theorem \ref{consistency}}
Although verification model $g$ and transformation $H$ can be unknown for our method, we can treat them as known for the proof of consistency of the posterior distribution of $(\mu_1,\sigma_1, \mu_2, \sigma_2)$ since given the labels and ranks, the posterior distributions of $(\mu_1, \sigma_1, \mu_2, \sigma_2)$ no longer depend on $g$ and $H$.

Let the set of all permutations of $\{1,2,...,N\}$ be denoted by $\Omega_N$. We want to show the existence of a function $h^*:\Omega_1 \times \Omega_2 \times ... \times \{ 0,1,2,3 \}^{\infty} \rightarrow \mathbb{R}^- \times \mathbb{R}^+ \times \mathbb{R}^+ \times \mathbb{R}^+$, such that $(\mu_1,\sigma_1, \mu_2, \sigma_2)=h^*(\bm{R}_N, N \geq 1, (L_1, L_2, ...))$, then in view of Doob's Theorem (Ghosal and Van der Vaart(2017) Theorem 6.9 and Proposition 6.10), the proof Theorem 2.4 will follow.

Let $1 \leq i_1 < ...< i_{N^*} \leq N$ be the collection of subject's indices whose label is verified, i.e., $L_{i_j}=0, 1$ or $2$, and $j=1,...,N^*$. From Lemma 2.3, 
\begin{align*}
f_Q(t|L \neq 3) = &(1-\lambda_{1(\mu_1, \sigma_1, \mu_2, \sigma_2, g)}^*-\lambda_{2(\mu_1, \sigma_1, \mu_2, \sigma_2, g)}^*)\phi^*_{(\mu_1, \sigma_1, g)}(t) + \lambda_{1(\mu_1, \sigma_1, \mu_2, \sigma_2, g)}^* \phi^*(t) \\
&+ \lambda_{2(\mu_1, \sigma_1, \mu_2, \sigma_2, g)}^*\phi^*_{(\mu_2, \sigma_2, g)}(t).
\end{align*} 
So the overall verified samples can be regarded as being generated independently from this mixture distribution disregarding the true disease status. So we have, 
\begin{align*}
Q_{i_j}\stackrel{\mathrm{i.i.d.}}{\sim} &(1-\lambda_{1(\mu_1, \sigma_1, \mu_2, \sigma_2, g)}^*-\lambda_{2(\mu_1, \sigma_1, \mu_2, \sigma_2, g)}^*)\Phi^*_{(\mu_1, \sigma_1, g)} + \lambda_{1(\mu_1, \sigma_1, \mu_2, \sigma_2, g)}^* \Phi^*_{(g)}\\ 
&+ \lambda_{2(\mu_1, \sigma_1, \mu_2, \sigma_2, g)}^*\Phi^*_{(\mu_2, \sigma_2, g)},
\end{align*}
where $\Phi^*_{(\mu_1, \sigma_1, g)}$, $\Phi^*_{(g)}$ and $\Phi^*_{(\mu_2, \sigma_2, g)}$ are the cumulative distribution functions of $\phi^*_{(\mu_1, \sigma_1, g)}$, $\phi^*_{(g)}$ and $\phi^*_{(\mu_2, \sigma_2, g)}$, respectively. Thus we have
\begin{align*}
U_j^*=&\{(1-\lambda_{1(\mu_1, \sigma_1, \mu_2, \sigma_2, g)}^*-\lambda_{2(\mu_1, \sigma_1, \mu_2, \sigma_2, g)}^*)\Phi^*_{(\mu_1, \sigma_1, g)} \\
&+ \lambda_{1(\mu_1, \sigma_1, \mu_2, \sigma_2, g)}^* \Phi^*_{(g)} + \lambda_{2(\mu_1, \sigma_1, \mu_2, \sigma_2, g)}^*\Phi^*_{(\mu_2, \sigma_2, g)}\}(Q_{i_j}) \stackrel{\mathrm{i.i.d.}}{\sim} \mathrm{Uniform}(0,1).
\end{align*}
Let ($R'_{N^*1}, ..., R'_{N^*N^*}$) and $(L'_{N^*1}, ..., L'_{N^*N^*})$ be the rank vector and labels of $(U_1^*, ..., U_{N^*}^*)$ respectively. According to Theorem $a$ on page 157 of H\'ajek, J. and \v Sid\'ak(1967), we have 
\begin{align*}
\mathrm{E} \Bigg(U_j^* - \dfrac{R'_{N^*_{i_j}}}{N^*+1} \Bigg)^2 & = \dfrac{1}{N^*} \sum ^{N^*}_{k=1}\mathrm{E}\Big[\Big(U_j^*-\dfrac{k}{N^*+1} \Big)^2| R'_{N^*_{i_j}}=k \Big] \\
&=\dfrac{1}{N^*} \sum ^{N^*}_{k=1} \dfrac{k(N^*-k+1)}{(N^*+1)^2(N^*+2)} < \dfrac{1}{N^*},
\end{align*}
so $\mathrm{E} \Bigg(U_j^* - \dfrac{R'_{N^*_{i_j}}}{N^*+1} \Bigg)^2 \rightarrow 0$ as $N \rightarrow \infty$. Therefore, there exists a subsequence \{$N^*_k$\} of $\{N^*\}$ such that for $j \geq 1$, $U_j^* = \lim_{k \rightarrow \infty} \dfrac{R'_{N^*_{k_{i_j}}}}{N^*_k+1}, a.s.$. Thus we can claim $U_j^*=h_j(\bm{R}_N, N \geq 1, L_1, L_2, ...)$ for some function $h_j:\Omega_1 \times \Omega_2 \times ... \times \{ 0,1,2,3 \}^{\infty}  \rightarrow [0,1]$.

Now given $\{Q_{i_j}: L_{i_j}=1\} \stackrel{\mathrm{i.i.d.}}{\sim} \Phi^*_{(g)}$, so that $\{U^*_j: L_{i_j}=1\} \stackrel {\mathrm{i.i.d.}} {\sim} V_{(\mu_1, \sigma_1, \mu_2,\sigma_2, g)}$, where $V_{(\mu_1, \sigma_1, \mu_2,\sigma_2, g)}$ is the distribution of 
\begin{align*}
\{(1-\lambda_{1(\mu_1, \sigma_1, \mu_2, \sigma_2, g)}^* -\lambda_{2(\mu_1, \sigma_1, \mu_2, \sigma_2, g)}^*)\Phi^*_{(\mu_1, \sigma_1, g)}   + \lambda_{1(\mu_1, \sigma_1, \mu_2, \sigma_2, g)}^* \Phi^*_{(g)}  \\
+  \lambda_{2(\mu_1, \sigma_1, \mu_2, \sigma_2, g)}^*\Phi^*_{(\mu_2, \sigma_2, g)}\}(\xi),
\end{align*} 
with  $\xi \sim \Phi^*_{(g)}$. 
Since $(U^*_j: L_{i_j}=1)$ are $\mathrm{i.i.d}$, $V_{(\mu_1, \sigma_1, \mu_2,\sigma_2, g)}$ is consistently estimable. So we only need to show that the family of $\{V_{(\mu_1, \sigma_1, \mu_2,\sigma_2, g)}: \mu_1<0, \sigma_1>0, \mu_2>0, \sigma_2>0\}$ is identifiable, i.e., if $V_{(\mu_1, \sigma_1, \mu_2,\sigma_2, g)}=V_{(\mu_1', \sigma_1', \mu_2',\sigma_2', g)}$, then $(\mu_1, \sigma_1, \mu_2, \sigma_2)=(\mu_1', \sigma_1', \mu_2', \sigma_2')$.

To prove that, first of all, we can show that $\Phi^*_{(\mu, \sigma, g)}$ and $\Phi^*_{(\mu', \sigma', g)}$ are linearly independent of each other given that $(\mu, \sigma) \neq (\mu', \sigma')$. Otherwise, there exists some $c_1, c_2 \in \mathbbm{R}$ such that $c_1 \Phi^*_{(\mu, \sigma, g)}(t) + c_2 \Phi^*_{(\mu', \sigma', g)}(t) = 0$ for all $t$, i.e., $c_1 \phi^*_{(\mu, \sigma, g)}(t) + c_2 \phi^*_{(\mu', \sigma', g)}(t) = 0$ for all $t$. This contradicts with the linear independence between $\Phi_{(\mu, \sigma)}$ and $\Phi_{(\mu', \sigma')}$. In addition, because $\mu_1 <0$ and $\mu_2>0$, so $\Phi^*_{(\mu_1, \sigma_1, g)}$ and $\Phi^*_{(\mu_2, \sigma_2, g)}$ are identifiable from each other. Thus we have $\lambda_{1(\mu_1, \sigma_1, \mu_2, \sigma_2, g)}^*=\lambda_{1(\mu_1', \sigma_1', \mu_2', \sigma_2', g)}^*$, $\lambda_{2(\mu_1, \sigma_1, \mu_2, \sigma_2, g)}^*=\lambda_{2(\mu_1', \sigma_1', \mu_2', \sigma_2', g)}^*$, $\Phi^*_{(\mu_1, \sigma_1, g)}=\Phi^*_{(\mu_1', \sigma_1', g)}$, $\Phi^*_{(\mu_2, \sigma_2, g)}=\Phi^*_{(\mu_2', \sigma_2', g)}$, which then implies that $(\mu_1, \sigma_1, \mu_2, \sigma_2)=(\mu_1', \sigma_1', \mu_2', \sigma_2')$.

Thus we can conclude that for some function $h$ of $(U_1, U_2, ..., L_1, L_2, ...)$, and consequently for a function $h^*$ of all ranks and labels,
\begin{equation*}
\begin{aligned}
(\mu_1, \sigma_1, \mu_2, \sigma_2, g) &= h(U_1, U_2, ...) \\
&= h(h_1(\bm{R}_N, N \geq 1, L_1, L_2, ...), h_2(\bm{R}_N, N \geq 1, L_1, L_2, ...), ...) \\
&= h^*(\bm{R}_N, N \geq 1, L_1, L_2, ...).
\end{aligned}
\end{equation*}
This justified the condition of Doob's theorem and hence theorem holds at $(\mu_{1}^*, \sigma_{1}*, \mu_{2}^*, \sigma_{2}^*) a.e. [\nu]$.
\end{proof}

\nocite{scurfield1996multiple}
\nocite{gu2014bayesian}
\nocite{greenwade93}
\nocite{nakas2014developments}
\nocite{nakas2004ordered}
\nocite{kang2013estimation}
\nocite{li2009nonparametric}
\nocite{inacio2011nonparametric}
\nocite{hsieh1996nonparametric}
\nocite{nze2015generalized}
\nocite{little2014statistical}
\nocite{chi2008receiver}
\nocite{alonzo2005assessing}
\nocite{ghosal2017fundamentals}
\nocite{duc2017bcrocsurface}
\nocite{xiong2006measuring}
\nocite{luo2013youden}
\nocite{duc2016bias}
\nocite{hajek1967theory}
\nocite{mossman1999three}
\nocite{youden1950index}

\bibliographystyle{plain}
\bibliography{trinorm_roc}

\end{document}